# PyFocus – a Python package for vectorial calculations of focused optical fields under realistic conditions. Application to toroidal foci.

Fernando Caprile[1,2], Luciano A. Masullo[1,2*], Fernando D. Stefani[1,2*]

[1]Centro de Investigaciones en Bionanociencias (CIBION), Consejo Nacional de Investigaciones Científicas y Técnicas (CONICET), Godoy Cruz 2390, C1425FQD, Ciudad Autónoma de Buenos Aires, Argentina

[2] Departamento de Física, Facultad de Ciencias Exactas y Naturales, Universidad de Buenos Aires, Güiraldes 2620, C1428EHA, Ciudad Autónoma de Buenos Aires, Argentina

* Corresponding authors. E-mail addresses: lmasullo@df.uba.ar and fernando.stefani@df.uba.ar

ABSTRACT

Focused optical fields are key to a multitude of applications involving light-matter interactions, such as optical microscopy, single-molecule spectroscopy, optical tweezers, lithography, or quantum coherent control. A detailed vectorial characterization of the focused optical fields that includes a description beyond the paraxial approximation is key to optimize technological performance as well as for the design of meaningful experiments and interpret properly their results. Here, we present PyFocus, an open-source Python software package to perform fully vectorial calculations of focused electromagnetic fields after modulation by an arbitrary phase mask and in the presence of a multilayer system. We provide a graphical user interface and high-level functions to easily integrate PyFocus into custom scripts. Furthermore, to demonstrate the potential of PyFocus, we apply it to extensively characterize the generation of toroidal foci with a high numerical aperture objective, as it is commonly done in super-resolution fluorescence microscopy methods such as STED or MINFLUX. We provide examples of the effects of different experimental factors such as polarization, aberrations, and misalignments of key optical elements. Finally, we present calculations of toroidal foci through an interface of different mediums, and, to our knowledge, the first calculations of toroidal foci generated in total internal reflection conditions.

PROGRAM SUMMARY

Program Title: PyFocus

Program Files doi: https://github.com/stefani-lab/ , https://pypi.org/project/PyCustomFocus/

User manual: https://pyfocus.readthedocs.io/

Licensing provisions: MIT

Programming language: Python 3.6

Nature of problem: Full vectorial calculation of focused of optical fields under realistic conditions, such as high numerical aperture, incident beams modulated in phase and/or intensity, and focusing through an interface or a multilayer system.

Solution method: PyFocus incorporates all the equations necessary to calculate the focused optical field and provides suitable integration methods. Setting up of the parameters, running calculations, plotting and saving results can be done by means of a graphical user interface or high-level functions.

Additional comments including restrictions and unusual features: In addition to the Python source code, an MS Windows executable is provided.

## 1. Introduction

Strongly focused optical fields are used in a multitude of technological applications and scientific research areas. Some examples include high-resolution light microscopy, lithography, optical tweezers, micromanipulation, cold ions trapping, single-molecule detection, quantum coherent control, and fluorescence nanoscopy among others.

Detailed knowledge of the focused optical fields in terms of their intensity, polarization, and phase distribution is key to achieve optimal performance in technological applications, as well as to design efficient experiments and properly interpret their results. The theory to compute focused optical fields under realistic experimental conditions is fully known. There is extensive literature covering the theory and analytical treatment of vectorial optical field equations[1,2]. Also, numerical solutions have been studied in many special cases[3,4]. However, either due to the lack of analytical solutions or the difficulty of the numerical calculations, one or more approximations are often used, namely paraxial approximation (small angles with respect to the optical axis), scalar approximation (polarization of light neglected) or perfect alignment and aberration-free cases are considered.

Here we present PyFocus, an open-source pure Python package for the fully vectorial calculation of focused optical fields after modulation by an arbitrary phase mask, delivering the intensity, phase and polarization of the resulting field. PyFocus also permits to take into account the propagation of the field from the phase mask to the focusing objective lens, as well as the calculation of focused fields through an interface or a multilayer system.

To illustrate the capabilities of PyFocus, we present calculations of:

i) focusing a beam with its phase modulated with a vortex-phase (VP) mask, as it is widely used in optical nanoscopy techniques such as STED[5,6], RESOLFT[7,8], and MINFLUX[9,10];

ii) focusing a beam near planar dielectric interface, which is of great importance for optical microscopy techniques using water or oil-immersion objective lenses, allowing the simulation of foci obtained by total internal reflection;

iii) the effect of common misalignments and optical aberrations on the VP-modulated focused beam.

PyFocus provides an application programming interface (API) for more advanced users but also a graphical user interface (GUI) that allows running simulations directly without the need for any further coding. The code is available at https://github.com/stefani-lab/ and can be installed directly as a Python package called PyCustomFocus (https://pypi.org/project/PyCustomFocus/). To make the package even easier to use we also provide an executable version that can be readily run in MS Windows even without installing a Python environment.

## 2. Electric field at the focus of an aplanatic lens

In an aberration-free optical system, the time-independent part of the electric field $E_f$ in a point $r$ near the geometric focus of a microscope objective can be calculated by adding the contributions of an angular distribution of electromagnetic plane waves traveling to the focus as [1,2]:

$$E_f = \frac{-ikf\,e^{-ikf}}{2\pi} \int_0^{2\pi} \int_0^{\alpha} E_0 e^{i\mathbf{k}\cdot\mathbf{r}} \sqrt{\cos(\theta)}\, \sin(\theta)\, d\theta\, d\phi' \qquad (1)$$

Where $f$ is the focal length of the objective, $\mathbf{k}$ is the wavevector, $\theta$ the angle of incidence and $E_0$ the complex amplitude of each plane wave traveling to the focus. The schematics in Figure 1a show the relevant parameters, coordinate systems used, and field components of a focusing wave with linear polarization. The spherical coordinates $(\theta, \phi')$ define the incident direction and starting point on the objective lens of the plane waves, while $\mathbf{r} = (\rho, \phi, z)$ defines the point of interest from the geometrical focus in cylindrical coordinates. Then, $\mathbf{k} \cdot \mathbf{r} = k(z\cos(\theta) + \rho\sin(\theta)\cos(\phi' - \phi))$. For an aplanatic lens, the radial position of each incident wave $(\rho')$ is given by the sine condition [1,2]: $\rho' = f\sin(\theta)$. $E_0$ can be calculated from the $s$ and $p$ components of the field incident on the lens $(E_i)$, given by $E_{i_s} = \mathbf{E}_i \cdot \hat{\boldsymbol{\phi}}'$, $E_{i_p} = \mathbf{E}_i \cdot \hat{\boldsymbol{\rho}}'$. As the field is focused, the $s$ component remains oriented in the same direction, while the $p$ component points in the direction given by the unit vector $\hat{\boldsymbol{\theta}}$ (Figure 1b). This results in: $\mathbf{E}_0 = [(\mathbf{E}_i \cdot \hat{\boldsymbol{\phi}}')\hat{\boldsymbol{\phi}}' + (\mathbf{E}_i \cdot \hat{\boldsymbol{\rho}}')\hat{\boldsymbol{\theta}}]$

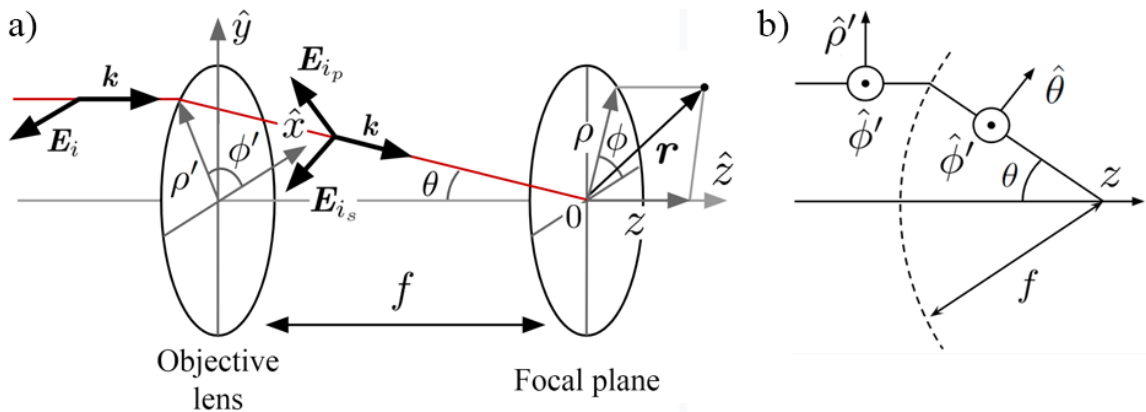

**Figure 1.** Geometry of the problem for the calculation of the electric field near a geometric focus. a) The incident field ($E_i$) is focused by an aplanatic objective lens. The coordinate system is taken such that the optical axis lies along the $\hat{z}$ direction. b) Geometric representation of the unit vectors $(\hat{\rho}', \hat{\phi}')$ for the incident field and $(\hat{\theta}', \hat{\phi}')$ for the focused field.

The obtained focused field can be expressed in cartesian components $\boldsymbol{E}_f = (E_{f_x}, E_{f_y}, E_{f_z})$ by applying the change of coordinates: $\hat{\boldsymbol{\rho}}' = cos(\phi')\hat{\boldsymbol{x}} + sin(\phi')\hat{\boldsymbol{y}}$, $\hat{\boldsymbol{\phi}}' = -sin(\phi')\hat{\boldsymbol{x}} + cos(\phi')\hat{\boldsymbol{y}}$ and $\hat{\boldsymbol{\theta}} = cos(\theta)cos(\phi')\hat{\boldsymbol{x}} + cos(\theta)sin(\phi')\hat{\boldsymbol{y}} - sin(\theta)\hat{\boldsymbol{z}}$. The total intensity ($I$) can then be calculated as the sum of each cartesian component squared: $I = E_{f_x}^2 + E_{f_y}^2 + E_{f_z}^2$.

*2.1 Focus of an elliptically polarized beam*

An elliptically polarized beam can be expressed as a linear combination of two orthogonally polarized beams, e.g. one polarized along $\hat{\boldsymbol{x}}$ and the other along $\hat{\boldsymbol{y}}$: $\boldsymbol{E}_i = E_i(c_x\hat{\boldsymbol{x}} + c_y\hat{\boldsymbol{y}})$, with $c_x$ and $c_y$ two complex constants such that $c_x^2 + c_y^2 = 1$. Then, the focused field produced by an elliptically polarized beam can be calculated as a linear combination of the field produced by focusing a beam with linear polarization along $\hat{\boldsymbol{x}}$ (called $\boldsymbol{E}_f^{(x)}$) and the field produced by focusing a beam with linear polarization along $\hat{\boldsymbol{y}}$ (called $\boldsymbol{E}_f^{(y)}$): $\boldsymbol{E}_f = c_x \boldsymbol{E}_f^{(x)} + c_y \boldsymbol{E}_f^{(y)}$.

$\boldsymbol{E}_f^{(x)}$ can be obtained by replacing $\boldsymbol{E}_i(\theta, \phi') = E_i(\theta, \phi')\hat{\boldsymbol{x}} = E_i(\theta, \phi')(1,0,0)$ in eq. 1, resulting in:

$$\boldsymbol{E}_f^{(x)}(\rho, \phi, z) = \frac{ikfe^{-ikf}}{2\pi} \int_0^{2\pi} \int_0^{\alpha} E_i(\theta, \phi') \begin{bmatrix} -cos(\theta) + (cos(\theta) - 1)sin^2(\phi') \\ (1 - cos(\theta))cos(\phi')sin(\phi') \\ sin(\theta)cos(\phi') \end{bmatrix} \cdot$$

$$\cdot e^{ik(zcos(\theta) + \rho sin(\theta)cos(\phi' - \phi))} \sqrt{cos(\theta)} sin(\theta) d\theta d\phi' \qquad (2)$$

One way to calculate $\boldsymbol{E}_f^{(y)}$ is by a coordinate rotation so that $\phi' \to \phi' - \pi/2$, giving as a result:

$$\boldsymbol{E}_f^{(y)}(\rho, \phi, z) = \frac{ikfe^{-ikf}}{2\pi} \int_0^{2\pi} \int_0^{\alpha} E_i(\theta, \phi') \begin{bmatrix} -(1 - cos(\theta))cos(\phi')sin(\phi') \\ -cos(\theta) + (cos(\theta) - 1)sin^2(\phi') \\ sin(\theta)cos(\phi') \end{bmatrix} \cdot$$

$$\cdot e^{ik(zcos(\theta) - \rho sin(\theta)sin(\phi' - \phi))} \sqrt{cos(\theta)} sin(\theta) d\theta d\phi' \qquad (3)$$

For most incident fields, these integrals do not have an analytical solution, hence requiring a two-dimensional (2D) numerical integration.

*2.2 Focus of a Gaussian beam*

An elliptically polarized Gaussian beam is described by

$$\boldsymbol{E}_i(\theta,\phi) = \boldsymbol{E}_G(\theta,\phi) = \boldsymbol{A}e^{-(\rho'/w_0)^2} \quad (4)$$

with $\boldsymbol{A} = A\left(c_x\hat{\boldsymbol{x}} + c_y\hat{\boldsymbol{y}}\right)$, and $w_0$ the radius at which the amplitude decreases to $e^{-1}$. Applying the sine condition, $\boldsymbol{E}_G(\theta,\phi) = \boldsymbol{A}e^{-(f\sin(\theta)/w_0)^2}$. It turns out that for $\boldsymbol{E}_G(\theta,\phi)$, the integrals over $\phi'$ of equations 2 and 3 have an analytical solution, as shown in Appendix A.1. In this case, only a one-dimensional (1D) numerical integration is needed.

*2.3 Focus of a beam modulated by a phase mask*

In this case, the field incident on the focusing lens corresponds to an initial field ($\boldsymbol{E}_e$) modulated by a phase mask of radius $R$. If the distance between the phase mask and the objective lens ($L$, Figure 2) is much longer than the wavelength, under the paraxial approximation $\boldsymbol{E}_i$ is given by Fraunhofer's diffraction formula:

$$\boldsymbol{E}_i(\rho',\phi') = \frac{-ik}{2\pi L}e^{ik(L+\rho'^2/2L)}\int_0^{2\pi}\int_0^R \boldsymbol{E}_e(\rho'',\phi'')e^{i\psi(\rho'',\phi'')}e^{ik/L\left(\rho''^2/2-\rho'\rho''\cos(\phi'-\phi'')\right)}\rho''d\rho''d\phi'' \quad (5)$$

where $\rho''$ and $\phi''$ are the polar coordinates for a point on the phase mask and $e^{i\psi(\rho'',\phi'')}$ is the phase term given by it on each point. In this calculation, it is assumed that no amplitude is transmitted outside the mask, and thus the integral goes from 0 to the phase mask radius ($R$). In general, this equation does not have an analytical solution and must be solved through a 2D numerical integration. The resolution at which $\boldsymbol{E}_i$ is calculated has a direct impact on the error of the integration for $\boldsymbol{E}_f$. For reasonable resolutions (1000 x 1000 pixels), this integration requires a computation time of tens of minutes in a personal computer.

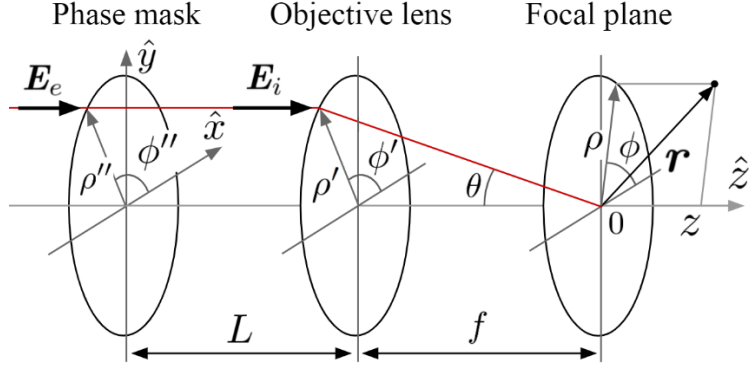

**Figure 2.** Geometrical scheme of the focusing of a beam modulated by a phase mask. An entrance field $E_e$ is modulated by a phase mask. The resulting field is propagated a distance $L$ to the objective lens obtaining the incident field $E_i$. The set of polar coordinates $(\rho'', \phi'')$ depicts a point located on the phase mask, while $(\rho', \phi')$ depict a point on the objective lens, and the spherical coordinates $(\rho, \phi, z)$ the point of interest $r$.

Alternatively, the propagation of the field from the phase mask to the objective can be neglected, and:

$$E_i = E_e e^{i\psi(\rho', \phi')} \quad (6)$$

*2.4 Focus of a beam modulated by a 0-2π vortex phase*

In the case of a *0-2π* vortex phase (VP) modulation, the phase term corresponds to $e^{i\psi(\rho'', \phi'')} = e^{i\pm\phi''}$, where the sign defines the handedness of the helicoidal modulation. In this work, the positive hand VP mask ($e^{i\psi(\rho'', \phi'')} = e^{i\phi''}$) will be considered. For a Gaussian $E_e$, equation 5 has an analytical solution on $\phi''$, and takes the following form after neglecting constant phase terms:

$$E_i(\rho', \phi') = \frac{kA}{L} e^{ik\rho'^2/2L} e^{i\phi'} \int_0^R e^{-(\rho''/w_0)^2} e^{ik\rho''^2/2L} J_1(k\rho''\rho'/L) \rho'' d\rho'' \quad (7)$$

If the propagation from the VP mask to the objective is neglected, the incident field simply results:

$$\boldsymbol{E}_i(\rho', \phi') = \boldsymbol{A}\, e^{-(\rho'/w_0)^2}\, e^{i\phi'} \qquad (8)$$

For both cases, since the topology of $\boldsymbol{E}_i$ in $\phi'$ is given by $e^{i\phi'}$, equations 2 and 3 have an analytical solution on $\phi'$, and the field near the focus can be calculated after a 1D numerical integration on $\rho'$. The resulting equations are shown in Appendix A.2.

*2.5 Focus through a multilayer system*

In experimental situations, it is often the case that the optical field is focused through layers of materials with different refraction indices. For example, in optical microscopy, water and oil immersion objectives are usually used, which imply the presence of at least one interface.

We will consider a multilayer system of neglectable total thickness in comparison to the focal distance ($f$), located at a distance $z_{int}$ from the focus, as schematically shown in Figure 3. The field previous and posterior to the multilayer is given by:

$$\boldsymbol{E}_f = \begin{cases} \boldsymbol{E}_{f_0} + \boldsymbol{E}_r & if\ z < z_{int} \\ \boldsymbol{E}_t & if\ z > z_{int} \end{cases} \qquad (9)$$

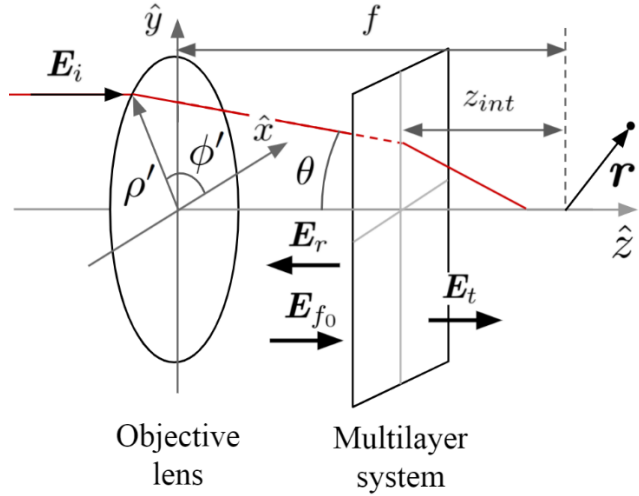

**Figure 3.** Geometry of the problem for the calculation of the electric field near a geometric focus with a multilayer system. An incident field $E_i$ is focused by an objective lens of focal distance $f$, producing the focused field $E_{f_0}$. This field incides on a multilayer system located at a distance $z_{int}$ from the geometrical focus. Negative values of $z_{int}$ denote a multilayer located previous to the focus. The multilayer causes the focused field to be refracted, producing a reflected field $E_r$ and a transmitted field $E_t$.

where $E_{f_0}$ is the focused field in the absence of the multilayer system. $E_r$ and $E_t$ are the fields reflected and transmitted by the multilayer system, which can be obtained by modifying equations 2 and 3 with the addition of the reflection and transmission coefficients for the $s$ and $p$ components of $E_i$, as shown in Appendix A.3. The overall reflection and transmission coefficients of the multilayer are obtained by solving Maxwell's equations through a Transfer Matrix Algorithm (TMA[20]). For the case of an interface between two materials, these correspond to the well-known Fresnel coefficients[2].

## 3. Overview of PyFocus

PyFocus provides high-level functions to configure and execute simulations of the focused field. While, in general, this requires a step of 2D numerical integration, for simulations without a multilayer system PyFocus includes the analytical solutions for a VP mask modulation and for an incident Gaussian beam without phase modulation to reduce the numerical computation to 1D integrals. For clarity, commands and syntax in PyFocus and Python are shown `in this format`. PyFocus also provides a graphical user interface (GUI) that allows an easy setup of the calculation parameters. Furthermore, we provide an executable version of PyFocus that can be run in a Windows environment with no need to install Python. In addition to the description of PyFocus contained in this article, a more detailed description of the code and instructions on how to install and use the package are provided at *https://pyfocus.readthedocs.io/* .

Examples of computing times correspond to a standard desktop PC running on 64 bits Windows 8.1, equipped with an AMD Athlon 200GE processor with a speed of 3.2GHz and 8Gb of RAM memory.

*3.1 Graphical user interface and calculation parameters*

PyFocus provides the `UI` class in `PyFocus.user_interface`. This allows opening the graphical user interface (GUI) upon creating an instance of this class (for example, `gui = PyFocus.user_interface.UI()`) and calling the `show()` function (for this example, `gui.show()`). The resulting GUI is shown in Figure 4. This can also be performed by running the *PyFocus.exe* executable file.

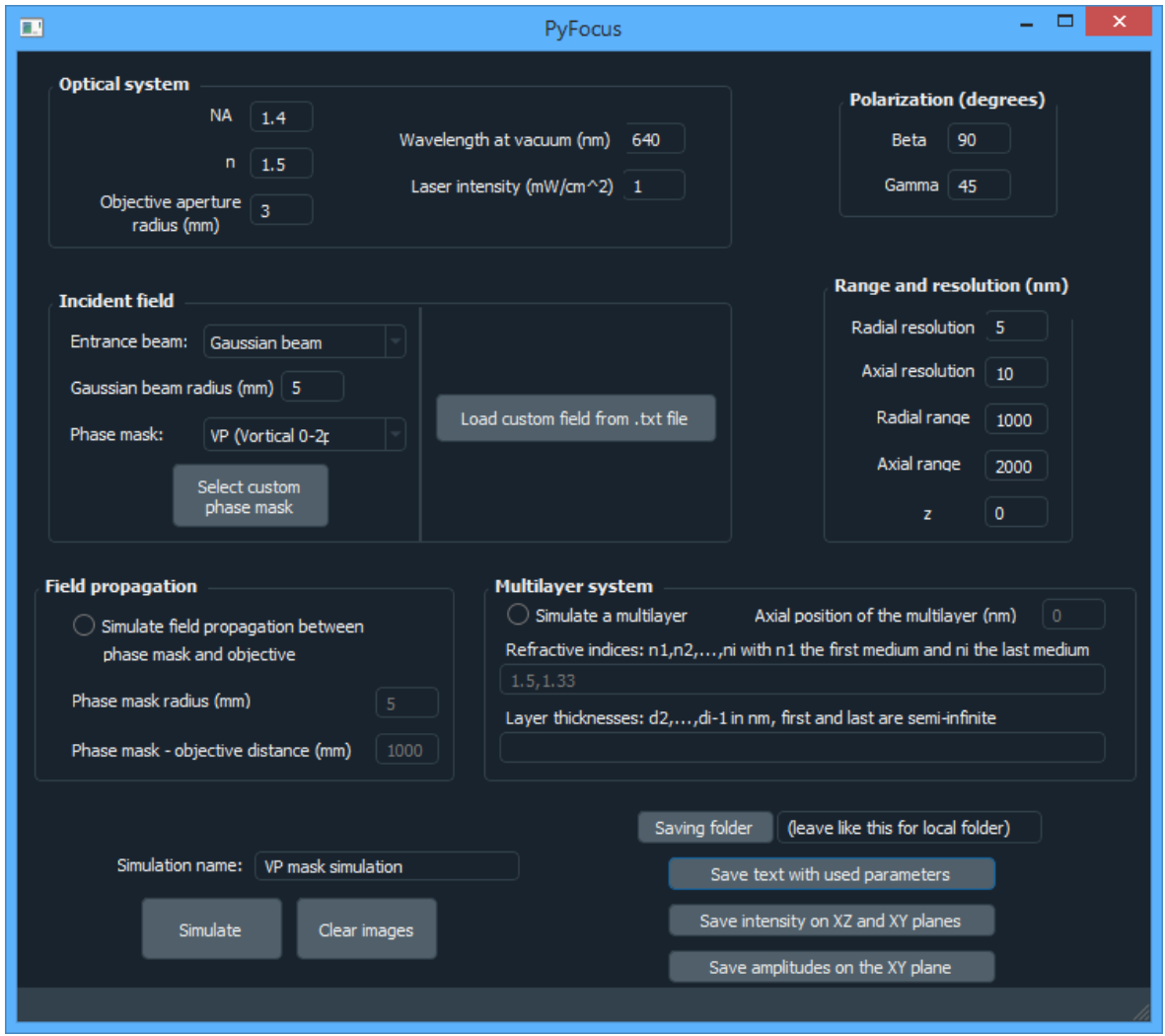

**Figure 4.** *PyFocus* graphical user interface (GUI).

The GUI facilitates the setting of the calculation parameters and the generation of output plots with the intensity of the focused field.

The parameters of the optical system and the incident field are schematically shown in Figure 5a. $R$ is the radius of the phase mask, $w_0$ the radius of the Gaussian beam (equation 4), and $L$ the distance between phase mask and objective lens. The parameters $h$ (objective lens aperture radius), $f$ (objective lens focal distance) and $n$ (refraction index of the focusing medium) are related through the definition of numerical aperture $NA = n\,sin(\alpha)$, where $\alpha$ is the maximum angle allowed by the aperture, defined as $\alpha = arcsin(h/f)$ following the sine condition.

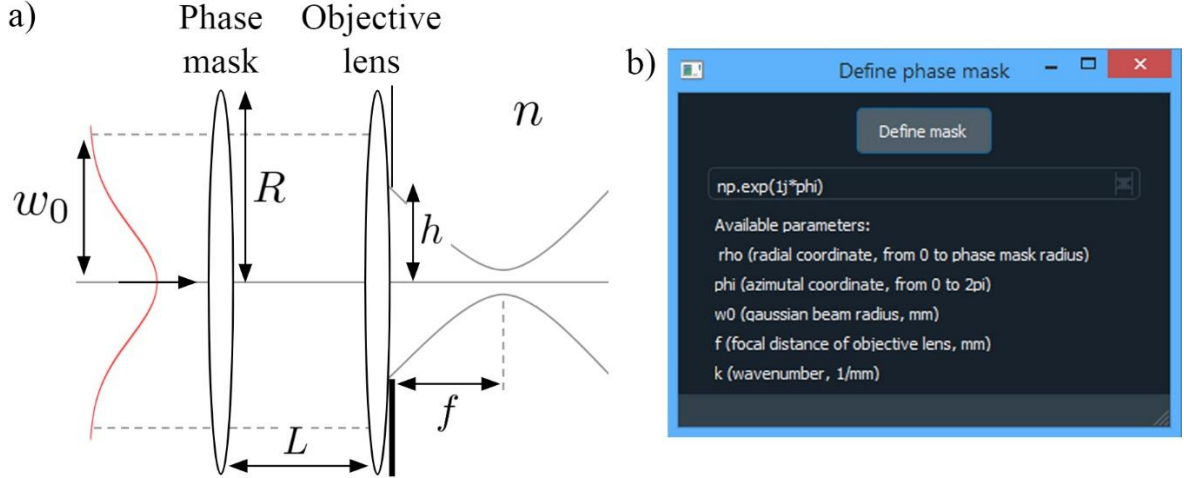

**Fig 5.** a) Experimental parameters that define a simulation. $w_0$: Gaussian beam's radius, $R$: radius of the phase mask, $L$: distance between phase mask and objective lens, $h$: aperture radius, $f$: focal distance. b) Dialog window that emerges when the "Custom mask" option is selected in the Phase Mask menu or when the "Select custom phase mask" button is pressed, allowing the definition of a custom phase mask by an analytical function.

The polarization of the beam at the entrance of the system is defined by two angular parameters ($\gamma$ and $\beta$): $\boldsymbol{E}_e = E_e(\cos(\gamma)\hat{\boldsymbol{x}} + \sin(\gamma)e^{i\beta}\hat{\boldsymbol{y}})$. In this way, $\gamma = 0°$ corresponds to linear polarization along $\hat{\boldsymbol{x}}$, and $\gamma = 45°$ and $\beta = 90°$ to a right circular polarization (from the point of view of the light-source), for example.

The range in which the focused field is calculated is given by the parameters *Radial range* and *Axial range*. The intensity of the focused field is calculated for points in the plane $z = 0$ ranging from $-\frac{Radial\ range}{2}$ to $\frac{Radial\ range}{2}$ in $\hat{x}$ and $\hat{y}$, and in the plane $y = 0$ for points ranging from $-\frac{Axial\ range}{2}$ to $\frac{Axial\ range}{2}$ in $\hat{z}$, and from $-\frac{\sqrt{2}\ Radial\ range}{2}$ to $\frac{\sqrt{2}\ Radial\ range}{2}$ in $\hat{x}$. The resolution of the focused field, i.e., the pixel size of the simulation is defined by the *Radial resolution* and *Axial resolution* parameters.

In the "Field propagation" box, a check-box enables choosing if the field propagation from the phase mask to the objective lens will be taken into account (equation 5) or not (equation 6). In case the propagation is considered, the phase mask radius $R$ and the separation distance between phase mask and objective lens $L$ must be provided.

In the "Multilayer system" box, a check-box enables choosing if the focus is produced through a multilayer or not. In case it is, the refractive indices and thicknesses of the layers must be provided in respective arrays. The (complex) refraction indices must be given as an array of the form $n_1, n_2, \ldots, n_i$, where $n_1$ is the index of the medium where the lens is located and $n_i$ is the index of the sample. The thickness of each layer must be given as an array of the same format omitting the first and last medium: $d_2, d_3, \ldots, d_{i-1}$, since these media are considered to be semi-infinite.

The incident field ($\boldsymbol{E}_i$) can be defined in the "Incident field" box in two ways: By loading a `.txt` file with the complex amplitude using the "Load custom field from .txt file" button (the requisites for this file are shown in Appendix B.1 and in the online documentation at *https://pyfocus.readthedocs.io/*) or by selecting an entrance beam ($\boldsymbol{E}_e$) profile in the "Entrance field" pop-up menu, allowing selection between a gaussian beam or uniform field, and selecting a phase mask in the "Phase mask" pop-up menu, allowing selection between a null mask, a VP mask and a custom mask given by the user. Upon selecting the use of a custom mask or by pressing the "Select custom phase mask" button, the dialog window shown in Figure 5b appears. This allows the definition of the phase mask ($e^{i\psi(\rho',\phi')}$) by an analytical function in Python syntax, with the real and complex part of the function depicting the amplitude and phase modulation of the mask respectively.

The "Simulate" button starts the calculations and shows a progress bar on the console. When finished, the output plots of the focused field are generated:

(I)     Intensity of the focused field on the $y = 0$ plane

(II)    Intensity of the focused field at the plane of constant axial position set by the parameter $z$

(III)   Intensity profile along the $y = 0$ axis of plot (II)

(IV-VI) Polarization and amplitude of each cartesian component on the plane defined by $z$.

All the generated plots can be cleared by pressing the "Clear images" button.

After finishing a simulation, the "Save text with used parameters" button generates a `.txt` file containing the parameters used for the simulation. The "Save intensity on XZ and XY planes" button generates a `.txt` file with the data of plots (I) and (II). The "Save amplitude on the XY plane" button generates a `.txt` file with the complex amplitude data of each cartesian component of plots (IV-VI). These files can be later loaded with the use of the function `numpy.loadtxt()`. The name used for these files can be set by the user in the "Simulation name" box, and the folder in which they are saved can selected by using the "Saving folder" button.

*3.3 Numerical integration methods*

1D numerical integrations are performed with the function `scipy.integrate.quad`. This function is based on an iterative method that ensures a minimal relative error ($< 1.5 \cdot 10^{-8}$). The number of iterations needed varies depending on the function to be integrated.

Solving 2D numerical integrations with `scipy.integrate.dblquad` (the two-dimensional version of `scipy.integrate.quad`) was found to be impractical because calculations with resolutions of only 100 x 100 pixels required computation times of up to several hours. Thus, 2D integrals were solved with a 2D trapezoidal integration algorithm, which has the advantage of solving the same integrals in times of only a few minutes. The disadvantage is that the error depends on the number of divisions used for the integration. This method was implemented with a custom-made code. In Appendix B.2 we provide further details for the appropriate number of divisions to be used.

*3.4 Use of PyFocus functions in custom scripts*

PyFocus simulations can also be performed through Python scripts using the functions from the `PyFocus.sim` module. The simulation parameters are set in the `parameters` array, given as a numpy (imported as `np`) array: `parameters = np.array((NA, n, h, w0, wavelength, gamma, beta, z, x_steps, z_steps, x_range, z_range, I0, L, R, ds, z_int, figure_name))`. Detailed description of all the input variables can be found in the documentation. The focal distance of the objective lens ($f$) is calculated by using the sine condition: $f = \frac{h\,n}{NA}$, and thus it is not defined in `parameters`.

The functions that calculate the focused field with no phase mask, a VP mask and a custom phase mask are `no_mask(propagation, multilayer, *parameters)`, `VP(propagation, multilayer, *parameters)`, and `custom(entrance_field, custom_mask, propagation, multilayer, *parameters, divisions_theta, divisions_phi, plot_Ei)` respectively. `propagation` and `multilayer` are Boolean variables to select if the propagation from the phase mask to the objective lens is to be simulated (by setting `propagation=True`) or neglected (`propagation=False`), and to select if a multilayer system is present for the simulation (`multilayer=True`) or not (`multilayer=False`). In the `custom` function, the entrance field ($E_e$) and the custom mask function ($e^{i\psi(\rho',\phi')}$) are given by the variables `entrance_field` and `custom_mask` respectively, which are explained in detail in the documentation, and the parameters `divisions_theta` and `divisions_phi` define the number of divisions for the 2D numerical integrations. `plot_Ei` is a Boolean variable that allows selecting whether to show figures with the intensity of $E_i$ (`plot_Ei=True`) or not (`plot_Ei=False`).

Each of these functions returns a tuple called `fields` including 6 arrays: 3 with the cartesian components of the focused field ($E_{f_x}, E_{f_y}, E_{f_z}$) on the plane $y = 0$ and 3 on the plane of axial position set by the parameter $z$. The obtained arrays can be plotted using the function `PyFocus.plot.plot_XZ_XY(*fields, x_range, z_range)`, where `x_range`, `z_range` correspond to the parameters *Radial range* and *Axial range* mentioned in the previous section. This function generates ten graphs, as shown in the examples of Figures 6 and 7. Four of them are shown in Figure 6: the intensity of the focused field on the plane y = 0 (Figure 6a), the intensity of the focused field on the plane $z = 0$ (Figure 6b), the intensity profile along the axis $y = 0$ of the plane $z = 0$ (Figure 6c), and a graph with the overlay of the total field intensity on the $z = 0$ plane together with the in-plane polarization (Figure 6d), the latter represented as ellipses with the axes being the $\hat{x}$ and $\hat{y}$ components ($E_{f_x}$ and $E_{f_y}$) and shown at points where the total intensity is greater than 15% of the maximum intensity. The other six graphs, shown in the example of Figure 7, are the intensity of each cartesian component ($E_{f_x}^2, E_{f_y}^2, E_{f_z}^2$) (Figures 7a-c) and their corresponding phases (Figures 7d-f) on the $z = 0$ plane.

In case of a multilayer system, the parameter `n` is the array with the refraction indices of each medium (`n=np.array(`$n_1, n_2, ..., n_i$`)`), and the parameter `ds` is the array with the thickness of each layer (`np.inf`: `ds=np.array(np.inf,`$d_2, d_3, ..., d_{i-1}$`,np.inf)`).

Code examples on how to use PyFocus functions are given in the documentation.

# 4 Examples using PyFocus. Analysis of toroidal foci.

In the following section, we show example calculations using PyFocus. Unless otherwise stated, the parameters used were $NA = 1.4$, $n = 1.5$, wavelength at vacuum $\lambda = 640$ nm, an entrance Gaussian beam of radius $w_0 = 5$ mm and intensity 1 mW/cm², $h = 3$ mm, $R = 5$ mm. To allow an easier comparation of the time required for the simulations, all figures were calculated with a resolution of 200 x 200 pixels.

*4.1 Focusing of a Gaussian beam with linear polarization along $\hat{x}$*

As a first example, we show the field obtained by focusing a Gaussian beam linearly polarized along $\hat{x}$ (Figure 6). This calculation can be done by selecting the "Null mask" option in the "Phase mask" pop-up menu of the GUI, or by using the `no_mask` function: `fields = sim.no_mask(propagation=False, multilayer=False, *parameters)`. After using the "Simulate" button in the GUI or by running the `plot` function: `plot.plot_XZ_XY(*fields,x_range,z_range)`, graphs showing the total intensity on the $y = 0$ plane (Figure 6a), the $z = 0$ plane (Figure 6b) and the $z = 0$, $y = 0$ axis (Figure 6c) are generated, as well as the polarization on the $z = 0$ plane (Figure 6d), the intensity of each cartesian component ($E_{fx}^2$, $E_{fy}^2$, $E_{fz}^2$) (Figures 7a-c) and their corresponding phases (Figures 7d-f) on the $z = 0$ plane. This simulation required an average computing time of 1 minute.

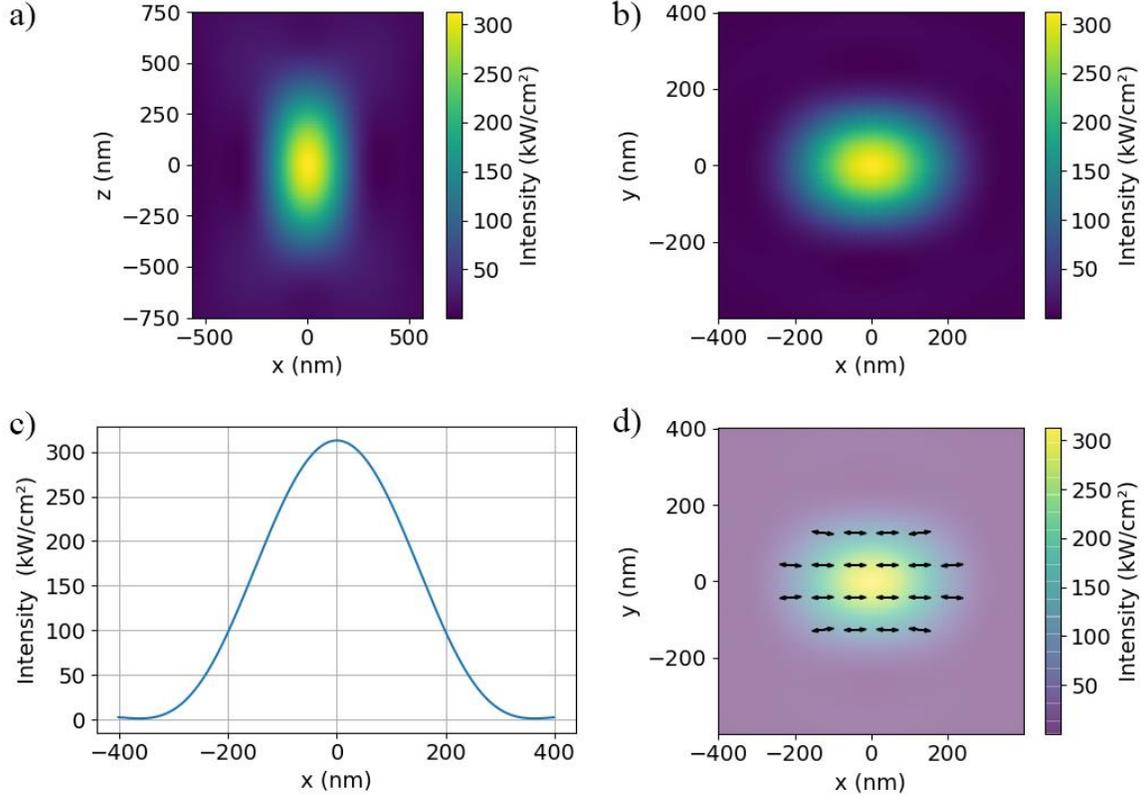

**Figure 6.** Total intensity ($I = E_{fx}^2 + E_{fy}^2 + E_{fz}^2$) of the field obtained by focusing a Gaussian beam with linear polarization along $\hat{x}$. a, b, c) Intensity along the $y = 0$ plane, the $z = 0$ plane and the $z = 0$, $y = 0$ axis respectively. d) Polarization on the $z = 0$ plane.

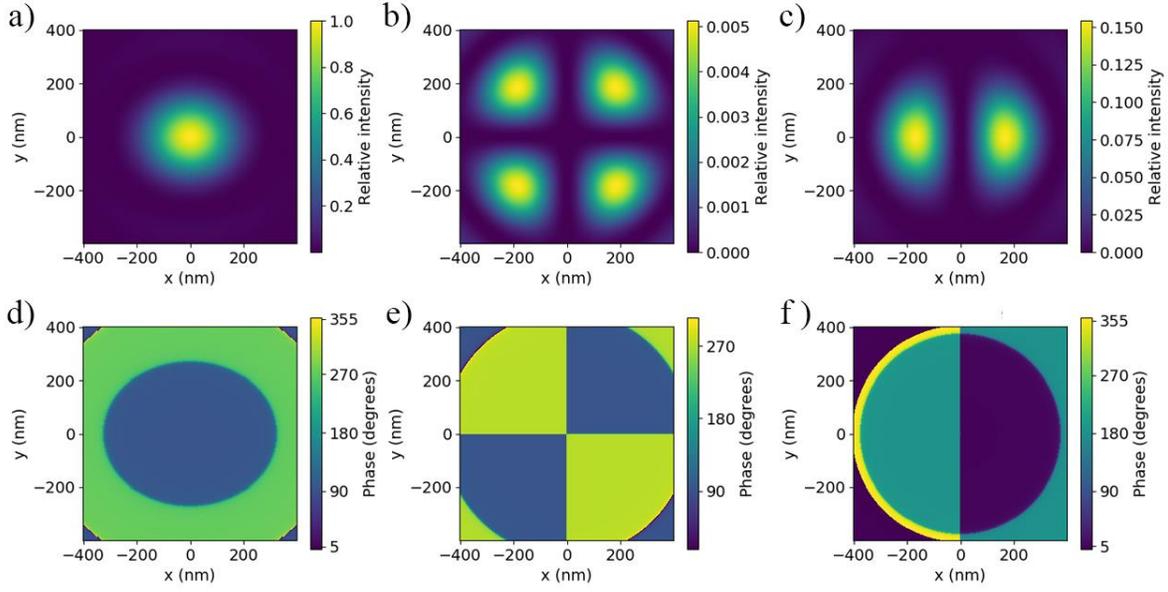

**Figure 7.** Intensity and phase of the cartesian components of the field obtained by focusing a Gaussian beam with linear polarization along $\hat{x}$. a, b, c) Intensity of the $\hat{x}, \hat{y}$ and $\hat{z}$ component ($E_{f_x}^2$, $E_{f_y}^2$ and $E_{f_z}^2$) respectively on the $z=0$ plane. d, e, f) Phase of the $\hat{x}, \hat{y}$ and $\hat{z}$ components respectively on the $z=0$ plane.

It is worth noting that due to the high *NA* of the objective, the focused field presents significant amplitudes in all cartesian components. However, the amplitudes of the $\hat{y}$ and $\hat{z}$ components are not high enough to modify the intensity pattern or the polarization appreciably.

*4.2 Toroidal foci variating the distance between the phase mask and the objective lens*

The field obtained by focusing a Gaussian beam modulated with a VP mask can be calculated by calling the `VP` function: `fields=sim.VP(propagation=False, multilayer=False, *parameters)` or by selecting the "VP (vortical 0-2pi)" option in the "Phase mask" pop-up menu of the GUI. These calculations use a positively handed VP mask. Therefore, in order to obtain a clean toroidal focus, an entrance field of right circular polarization (`gamma=45`, `beta=90`) should be used. If left circular polarization is used with this hand of VP, the resulting $\hat{z}$ component of the focused field has an intensity maximum at the center, which strongly decreases the contrast of the intensity minimum at the center of the pattern. An example of a simulation with an incident beam of left circular polarization is shown in Appendix A.4.

Figures 8a and 8b show the intensity distribution of the focused field in the $y = 0$ and $z = 0$ planes, respectively, where the toroidal shape can be observed. Figure 8c shows an intensity profile along $y = 0$, where it can be seen that the central minimum has zero intensity. The intensity and polarization in the $z = 0$ plane are shown together in Figure 8d.

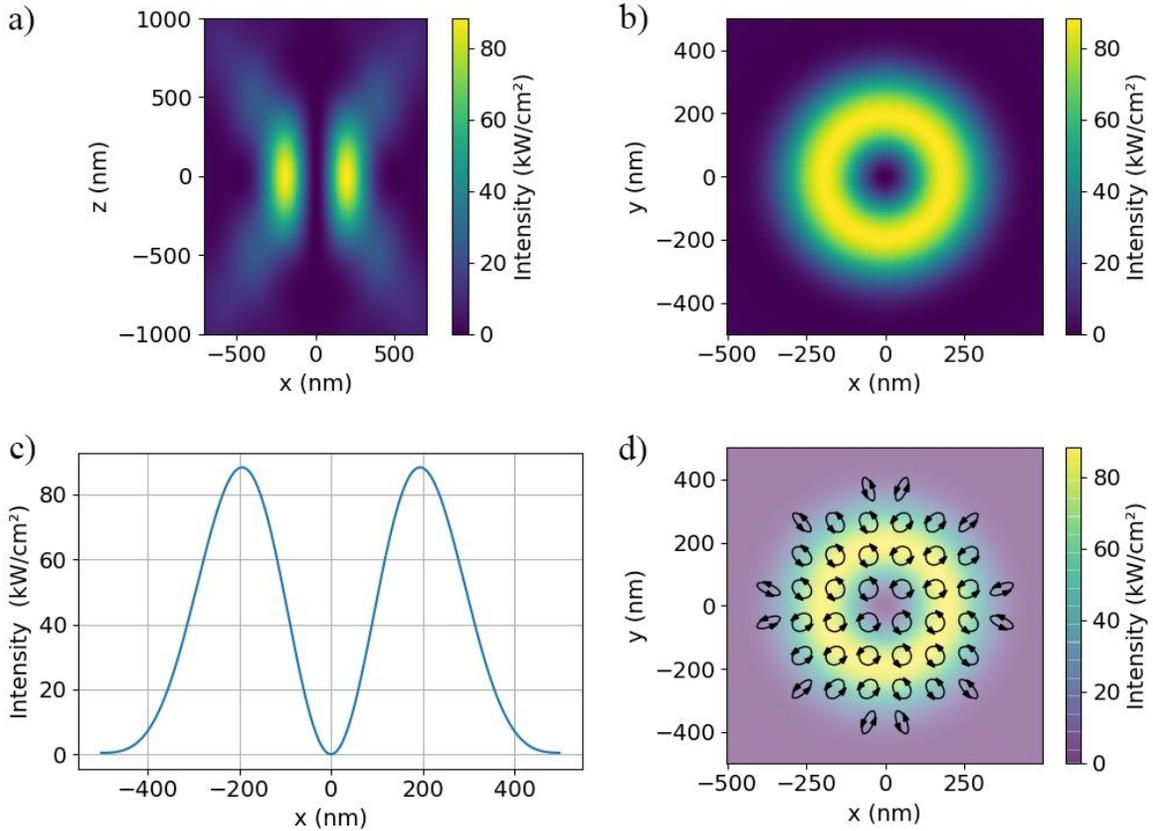

**Figure 8.** Intensity obtained by focusing a Gaussian beam of right circular polarization modulated by a VP mask. a, b, c) Intensity along the $y = 0$ plane, the $z = 0$ plane and the $z = 0$, $y = 0$ axis respectively. d) Polarization map on the focal plane.

The influence of the propagation of the incident field was studied by simulating the foci obtained for various distances of propagation (`L`) and setting the `propagation` parameter to `True`, which can also be done by selecting this option in the GUI. This way, running the `VP` function generates the graphs of $\boldsymbol{E}_i$ for distances of 0.2 m, 2 m, 10 m and 50 m shown in Figures 9a and 9b. Figure 9c shows the corresponding focused fields. The calculation of the propagated field required times ranging from 1 to 10 minutes for a resolution of 1000 x 1000 pixels (longer times required for smaller values of L) and 10 minutes for the calculation of the focused field.

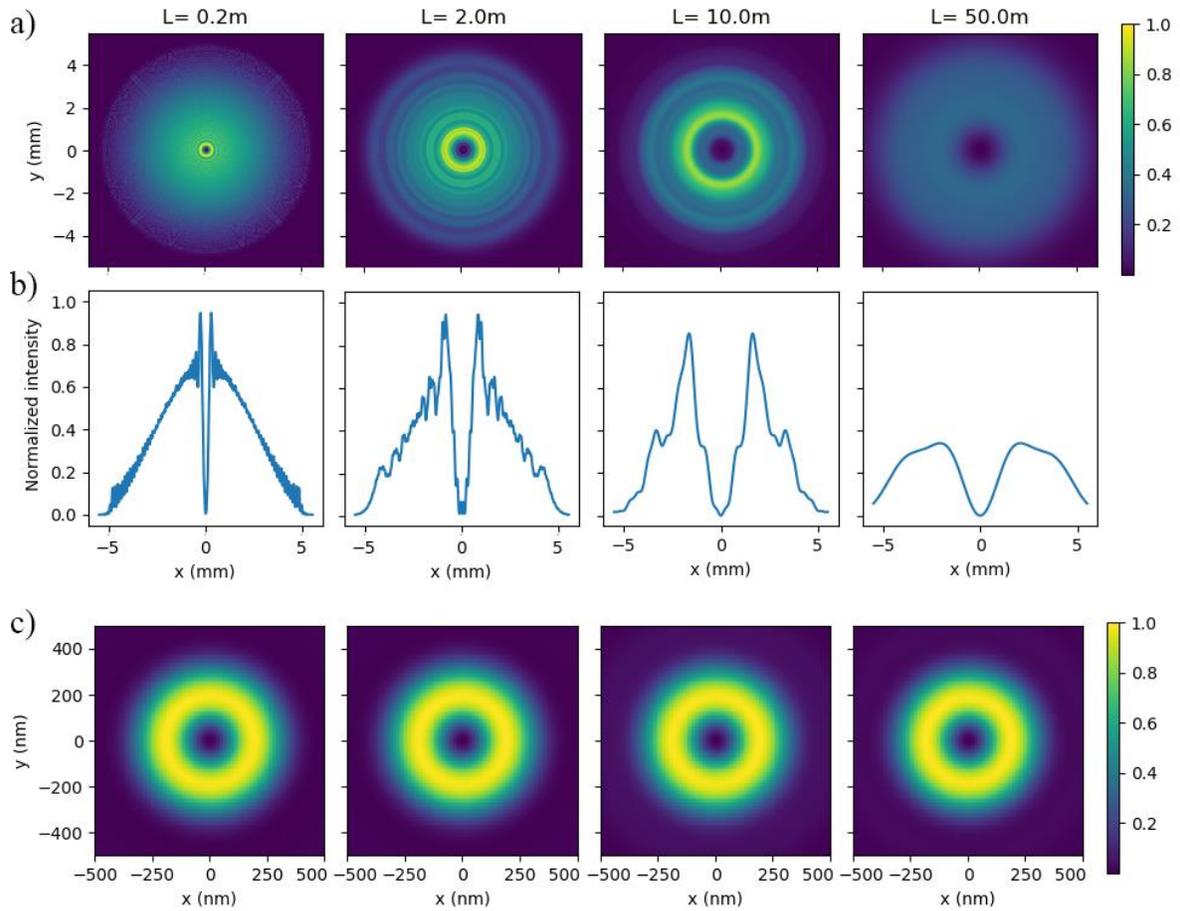

**Figure 9.** a) Normalized intensity of the field incident on the objective lens for various distances of propagation ($L$) between the phase mask and the lens. b) Profiles of a) along $y = 0$. The profiles are all normalized to the maximum intensity of the field for $L = 0.2$ m. c) Normalized focused field intensity distribution at $z = 0$ for the values of $L$ shown in a) and b).

Remarkably, although the shape of the incident beam is modified appreciably by varying $L$, the focused fields have no appreciable differences. The reason for this is that the phase term provided by the VP mask ($e^{i\phi}$) is still present in all the incident fields, regardless of their intensity spatial distribution. Because the propagation does not affect significantly the focused field of toroidal foci, `propagation = False` will be the default setting in the next simulations.

*4.3 Toroidal foci variating the Gaussian beam's radius*

PyFocus allows to easily analyze the foci obtained with different radius of the incident beam (w0). We parametrize these calculations in terms of the filling factor $F = \frac{w_0}{h}$, with $h$ (h) being the radius of objective aperture. A sufficiently large $F$ is equivalent to illuminate the objective lens with uniform intensity. In this case, the full $NA$ of the objective is exploited. Reducing $F$ is equivalent to decrease the effective $NA$ of the objective lens.

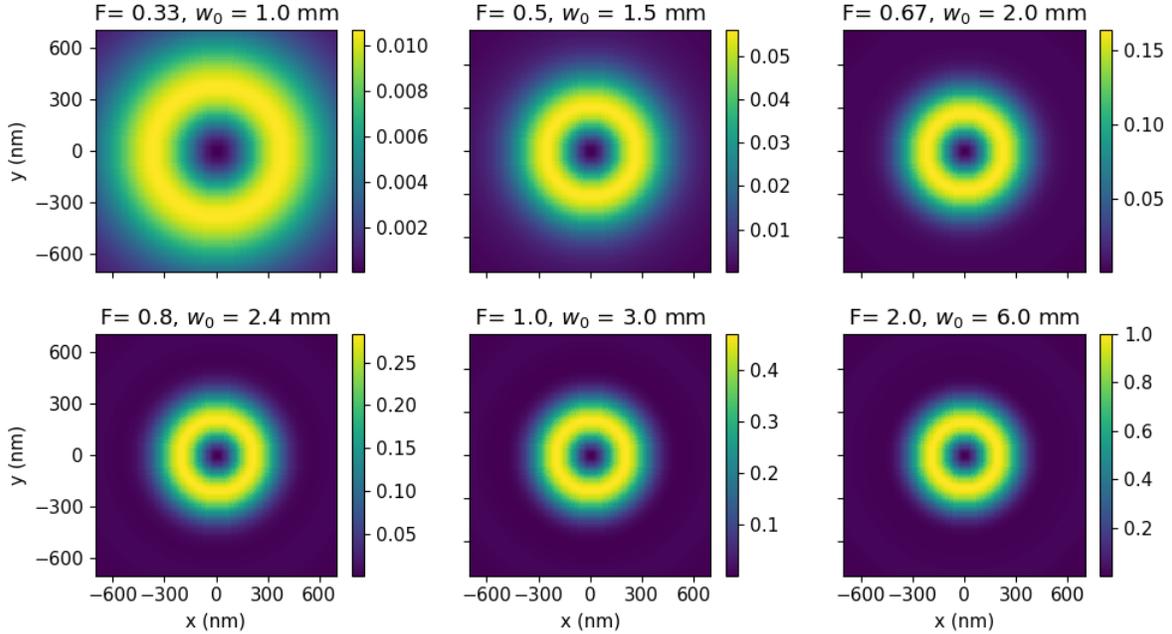

**Figure 10.** Intensity profile on the $z = 0$ plane of the toroidal foci obtained while variating the radius of the Gaussian beam ($w_0$), parametrized in terms of the filling factor $F = \frac{w_0}{h}$. All simulations were normalized with respect to the maximum intensity.

In Figure 10, toroidal foci calculated with various filling factors ranging from 0.33 to 2.0 are shown on the $z = 0$ plane. As $F$ decreases, the size of the foci increases. For example, for $F = 0.5$ the diameter of the toroidal focus (separation distance between points of maximum intensity across the central cero) is 508 nm, and reduces to 428 nm for F= 1.0, and to 392 nm for F= 2.0. The diameter obtained using uniform illumination (i.e., $F \to inf$) is 380 nm (not shown; calculated using $E_i(\rho', \phi') = e^{i\phi'}$).

*4.4 Custom masks. Simulation of misalignments*

PyFocus enables the calculation of focused fields modulated in phase by arbitrary masks. Together with modulations of amplitude, which can be introduced trivially, engineered focused fields can be calculated with full versatility. One application of these capacity is the simulation of aberrations due to the misalignment of the optical setup. For example, the effect of a laterally displaced or an inclined entrance beam can be simulated.

An off-axis incident field can be modeled by performing a shift of the center of coordinates as schematically shown in Figure 11a, where the center of coordinates $O$ is shifted a distance $d$ along the $\hat{x}$ axis. Then, any point of $\boldsymbol{E}_i$ defined by $(\rho', \phi')$ in the original coordinates system is now defined by $(\tilde{\rho}, \tilde{\phi})$:

$$\tilde{\rho} = \sqrt{\rho'^2 + d^2 - 2\rho' d \cos(\phi')}$$
$$\tilde{\phi} = \arctan\left(\frac{\rho' \sin(\phi')}{\rho' \cos(\phi') - d}\right) \tag{10}$$

A similar shift of coordinates can be applied to the phase mask or not, depending on the misalignment situation to be simulated. For example, Figure 11b shows a situation where both a Gaussian incident beam and the VP mask are displaced the same distance with respect to the objective. In this case, the incident field is $\boldsymbol{E}_i(\rho', \phi') = \boldsymbol{E}_G\bigl(\tilde{\rho}(\rho', \phi')\bigr) e^{i\tilde{\phi}(\rho', \phi')}$.

This incident field was simulated by using the function `custom`: `fields=sim.custom entrance_field, custom_mask , propagation=False, multilayer=False, *parameters, divisions_theta, divisions_phi, plot_Ei)`. The variables `entrance_field` and `custom_mask` are functions defining the terms $\boldsymbol{E}_e$ and $e^{i\tilde{\phi}(\rho', \phi')}$ respectively, and must be given as a function of five parameters: `rho`, `phi`, `w0`, `f` and `k`, which correspond to the variables $\rho'$, $\phi'$, $w_0$, $f$ and $k = \frac{2\pi}{\lambda}$ respectively. For a displacement `d`, the entrance field was then defined as: `entrance_field = lambda rho,phi,w0,f,k: np.exp(-(rho**2+d**2-2*rho*d*np.cos(phi))/w0**2)` and the custom mask as `custom_mask = lambda rho,phi,w0,f,k: np.exp(1j*np.arctan2(rho*np.sin(phi), rho*np.cos(phi)-d))`. To verify that the displacement was correctly implemented, the option `plot_Ei=True` can be used to obtain graphs of the total intensity of $E_i$, as shown in Figures 11c and 11d, and of the intensity and phase of the $\hat{x}$ and $\hat{y}$ components ($E_{i_x}$, $E_{i_y}$), as shown in Figures 11e-h.

`divisions_theta` and `divisions_phi` were both set to 200 because this resolution was found to be sufficient for accurate calculations with computation times of a few tens of minutes; see Appendix B.2 for a more detailed discussion.

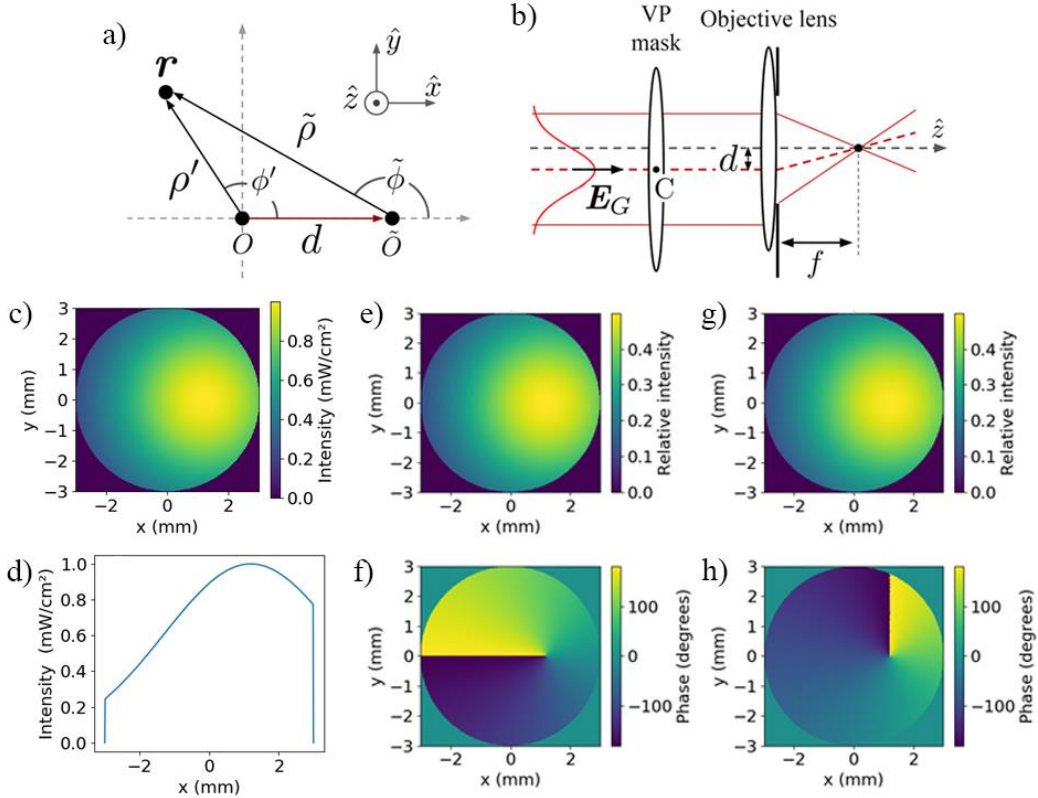

**Figure 11.** Simulation of an off-axis incident field. a) Schematic of the change of coordinates from $O$ to the displaced center of coordinates $\tilde{O}$. b) Scheme of a situation where an incident Gaussian beam and a VP mask are both displaced off-axis by the same distance $d$ along $\hat{x}$. The center of the phase mask (C) is shown. c) Example calculation of the intensity of $E_i$ at the entrance plane of the objective lens for $d = 1$ mm. d) Intensity profile of c) along the $z = 0, y = 0$ axis. e, f) Intensity and phase of $E_{i_x}$. g, h) Intensity and phase of $E_{i_y}$.

Figure 12 displays the obtained focused fields when the incident Gaussian beam and the phase mask are displaced together in different degree, parametrized by $D = \frac{d}{h}$. Each simulation required an average computing time of 4 minutes.

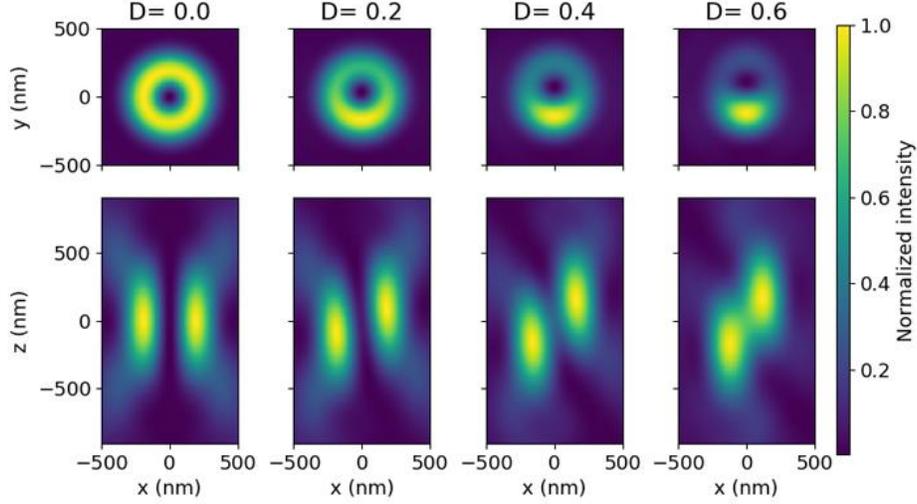

**Figure 12.** Simulation of an off-axis displacement of the incident field in the formation of a toroidal focus. Intensity obtained for a displaced Gaussian beam and VP mask along $\hat{x}$ described in terms of the parameter $D = \frac{d}{h}$. The intensity for each field was normalized individually.

The obtained foci are notably distorted from the toroidal shape as D increases. We note that the level of distortion also depends on the radius of the Gaussian beam ($w_0$). For example, if $w_0 \to \infty$, the amplitude of the incident field remains unchanged upon misalignments and the only aberration consists in the displacement of the vortical phase term.

An incident beam with an inclination (tilt) with respect to the optical axis ($\hat{z}$) can be modeled by a term that accounts for the difference in the optical path of each point: $e^{ik\Delta(\rho',\phi')}$. Figure 13a shows a schematic of an incident beam tilted an angle $\alpha$ on the $y = 0$ plane. In this case the term results: $e^{ik\Delta(\rho',\phi')} = e^{ik\rho' \sin(\alpha)\cos(\phi')}$. Following an optical ray tracing, it can be predicted that the focus will be displaced a distance $d = f \tan(\alpha)$ along $\hat{x}$, as schematically shown in Figure 13b.

The incident field on the lens for a tilted Gaussian beam modulated by a VP mask results: $\boldsymbol{E}_i(\rho',\phi') = \boldsymbol{E}_G(\rho')\, e^{i\phi'} e^{ik\rho' \sin(\alpha)\cos(\phi')}$, which was simulated by using the `custom` function and setting `entrance_field = lambda rho,phi,w0,f,k: np.exp(-(rho/w0)**2)` and `custom_mask = lambda rho,phi,w0,f,k: np.exp(1j*(phi+k*np.sin(alpha)*rho*np.cos(phi)))`, with `alpha` the angle of tilt $\alpha$.

Figure 13c show example simulations of foci obtained with rather small tilt angles $\alpha$ of 0.0009°, 0.0018° and 0.0027°, which correspond to displacements of 50 nm, 100 nm, and 150 nm, respectively. Each simulation required an average computing time of 4 minutes. These simulations can also be done in the GUI by selecting the option "Custom mask" in the "Phase mask" pop-up menu and setting the custom mask function as `np.exp(1j*(phi+k*np.sin(alpha)*rho*np.cos(phi)))`.

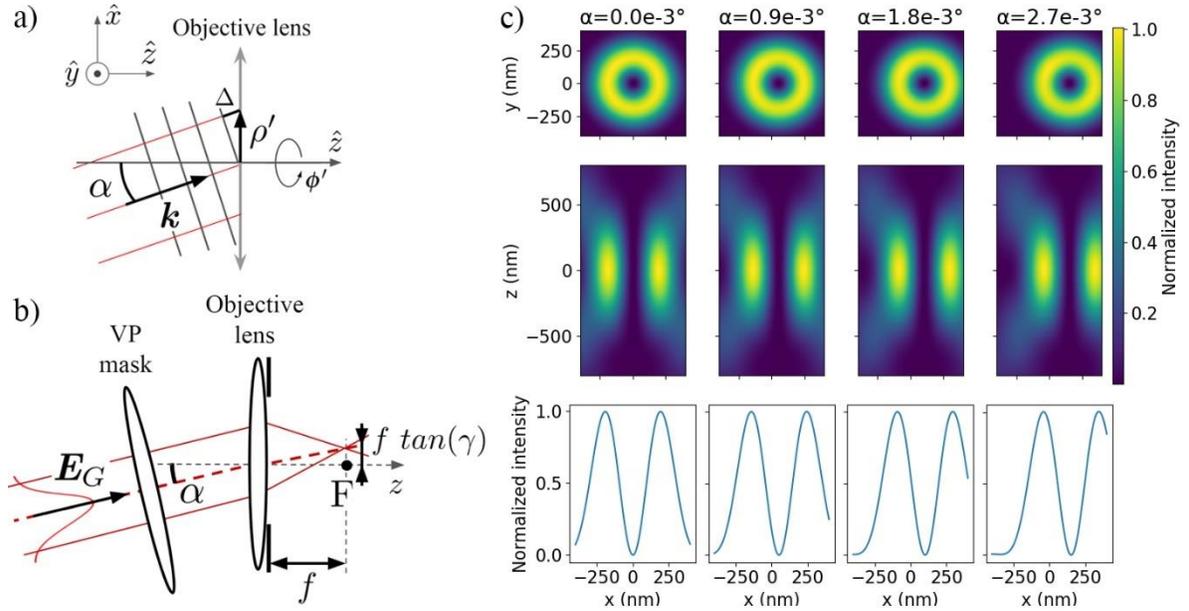

**Figure 13.** Toroidal foci obtained by a tilted incident beam. a) Geometry of the problem, the wavevector $\boldsymbol{k}$ of the incident field forms an angle $\alpha$ with the optical axis ($\hat{z}$). The plane of rotation is selected as the $y = 0$ plane. b) Scheme of the focusing of a tilted incident beam modulated by a VP mask. The geometrical focus of the objective lens (F) is shown. c) Toroidal foci obtained for angles of inclination that produce displacements of 50 nm, 100 nm and 150 nm.

For the simulated small angles, the toroidal shape is maintained without any appreciable differences, with the only effect being the displacement of the pattern as expected. We note that for larger values of $\alpha$, in addition to a larger displacement, an inclination of the focused field is produced.

*4.5 Custom masks to simulate aberrations with Zernike polynomials*

Optical aberrations can be modeled with the use of Zernike polynomials[11]. PyFocus allows to easily simulate the effect of different aberrations by using the analytical description of the polynomials as a custom mask. We show the effect of these aberrations on toroidal foci by using as custom mask $e^{i\psi(\rho',\phi')} = e^{i(\phi' + A_\Phi \Phi(\rho',\phi'))}$, with $\Phi(\rho',\phi')$ a Zernike polynomial and $A_\Phi$ the amplitude for that polynomial. The lower order normalized Zernike polynomials are: Astigmatism: $\Phi(r,\phi') = Z_2^{-2} = \sqrt{6}\, r^2 sin(2\phi')$, Spherical: $\Phi(r,\phi') = Z_2^0 = \sqrt{3}\, 2r^2$, Coma: $\Phi(r,\phi') = Z_3^{-1} = \sqrt{8}\, (3r^2 - 2)r\, sin(\phi')$ and Trefoil: $\Phi(r,\phi') = Z_3^{-3} = \sqrt{8}\, r^3 sin(3\phi')$, with $r = \frac{\rho'}{h}$. The obtained foci for each of these functions with an amplitude of $A_\Phi = 1.5$ rad is shown in Figure 14. Each of these simulations required an average computing time of 4 minutes.

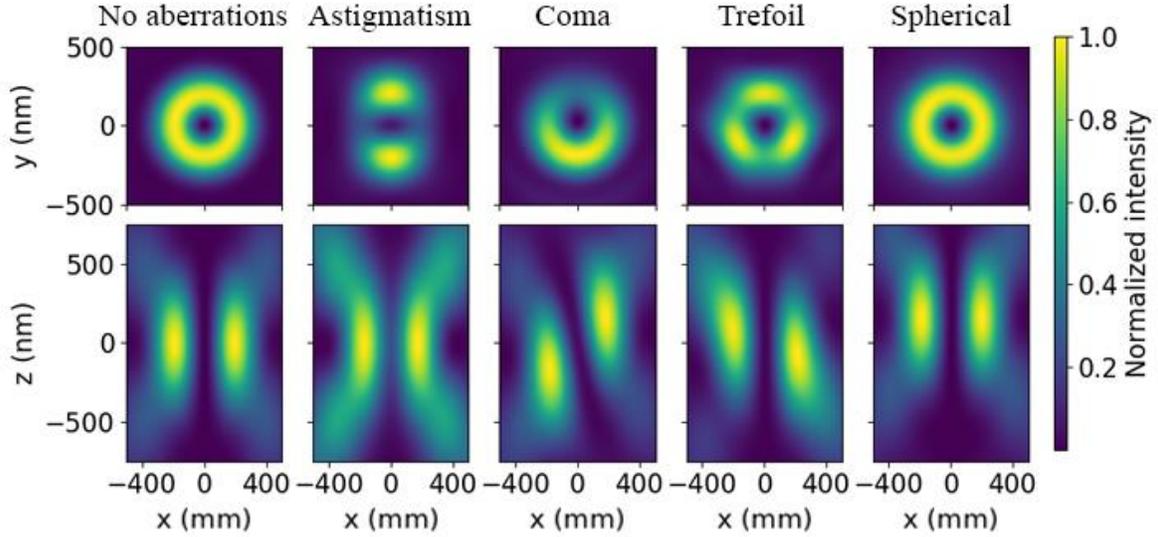

**Figure 14.** Simulation of the effect of aberrations in toroidal foci with the use of Zernike polynomials with an amplitude of $A_\Phi = 1.5$. Intensity is normalized individually for each focus.

## 5. Fields focused through an interface or a multilayer

Optical systems for fluorescence microscopy often use an interface of mediums (for example glass-water) in the focusing path, resulting in a change in the refractive index near the focal plane. There are also applications (for example in nanophotonics and plasmonics) in which a multilayer system, i.e., composed of multiple mediums with different refraction indexes, is of interest. PyFocus allows the calculation of focusing fields through a multilayer by setting the `multilayer` parameter to `True`, or by selecting this option with the "Simulate a multilayer" check box in the GUI.

*5.1 Toroidal foci through a glass-water interface*

First, we show a simulation of the interface between the immersion medium of the objective and the aqueous medium of the sample, which is of great importance for optical microscopy techniques using oil-immersion objective lenses. The refraction indices are considered to be $n_1 = 1.5$ for glass and $n_2 = 1.33$ for water. To perform this simulation using the GUI, the "Refraction indices" box must be set to: 1.5, 1.33 and the "Layer thicknesses" box must be left empty, since the first and last layers are considered semi-infinite. If using the functions provided by PyFocus, the parameters `n`, `ds` and `z_int` inside the `parameters` array must be set to `n=np.array((1.5,1.33))`, `ds=np.array((np.inf,np.inf))` and for an interface located at the focal plane: `z_int=0`. A toroidal foci with the use of a VP mask with a glass-water interface located the focal plane was then calculated by using the `VP` function: `fields=sim.VP(propagation=False, multilayer=True, *parameters)`. In Figure 15 we show a comparison between this field and the focus obtained without the interface. The simulation with the interface required an average computing time of 12 minutes.

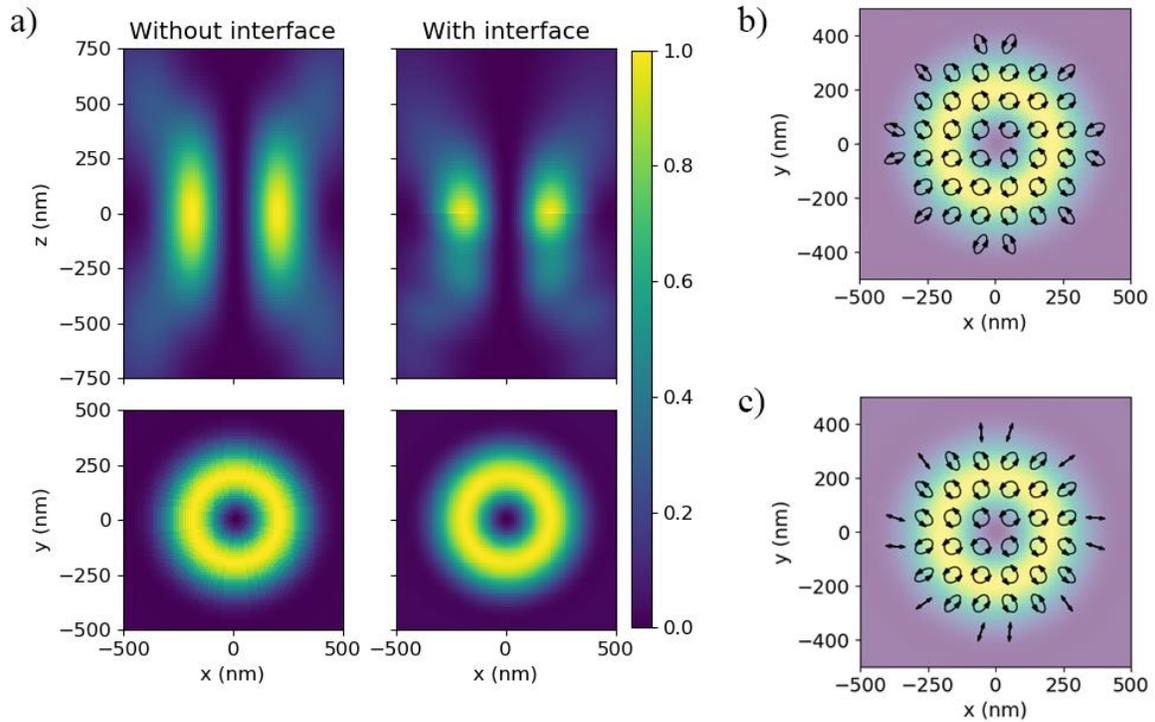

**Figure 15.** Comparison of a toroidal focus in water to a toroidal focus with a glass-water interface at the focal plane. a) Intensity distributions on the Y = 0 and Z = 5 nm planes. b, c) Intensity and polarization at the focal plane (Z = 5 nm) of the cases with and without the interface, respectively.

*5.2 Foci in total internal reflection for a glass-water interface*

Total internal reflection fluorescence (TIRF) microscopy [12] is a widely used technique to visualize structures close to an interface. TIRF provides an improved signal-to-background ratio, limits the detection to molecules located in proximity to the interface and, moreover, it enables retrieving the axial position of the single fluorophores[13]. Combining these properties of TIRF with focused beams have the potential to deliver synergistic solutions for several applications, as it has been demonstrated for fluorescence microscopy in pioneer work[14,15].

One way to generate total internal reflection (TIR) foci is by using objectives with $NA > 1$ and obstructing the fraction of the incident field that is focused with an inclination angle lower than the critical angle of incidence of the interface $\theta_c$. We first calculated TIR foci using a Gaussian incident field multiplied with an obstruction function called $g_{TIRF}$:

$$\boldsymbol{E}_i(\rho', \phi') = \boldsymbol{E}_G(\rho') g_{TIRF} = \begin{cases} \boldsymbol{E}_G(\rho') & \text{if } \rho' > f\sin(\theta_c) \\ 0 & \text{if } \rho' < f\sin(\theta_c) \end{cases} \quad (11)$$

Where $f$ is the focal distance of the objective lens. Using the `custom` function of PyFocus, this field was simulated by defining `entrance_field=np.exp(-(rho/w0)**2)` and `custom_mask` as:

```
theta_crit=np.arcsin(n[1]/n[0])
def custom_mask(rho,phi,w0,f,k):
    if rho>f*np.sin(theta_crit):
        return 1
    else:
        return 0
```

With `n` an array with the refraction indices of glass and water: `n=np.array(`$n_1, n_2$`)`.

Figures 16 a-d show the intensity distributions of TIR foci for an interface located at the focal plane, with an incident beam linearly polarized along $\hat{x}$ (Figures 16a and 16c) and right-hand circularly polarized (Figures 16b and 16d), and a comparison of their intensities on the $x = 0$, $y = 0$ axis (Figure 16e) and the $z = 5$ nm, $y = 0$ axis (Figure 16f). Each simulation required an average computing time of 12 minutes.

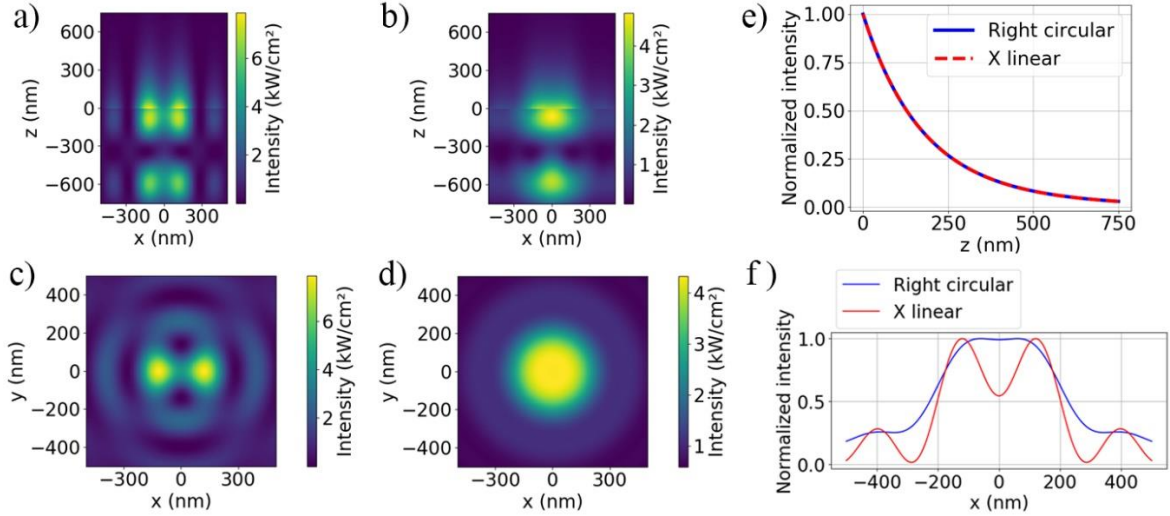

**Figure 16.** Total internal reflection foci on a glass-water interface located at the focal plane. a, c) Intensity distribution on the $y = 0$ plane and the $z = 5$ nm plane respectively of a TIR focus produced with incident beam with linear polarization along $\hat{x}$. b, d) Intensity distribution on the $y = 0$ plane and the $z = 5$ nm plane respectively of a TIR focus produced with an incident beam with right circular polarization. e, f) Intensity profiles along the $x = 0$, $y = 0$ axis and the $z = 5$ nm, $y = 0$ axis respectively of the foci shown in (a-d). Intensity is normalized individually for each profile.

The intensity distributions of the TIR foci in the focal plane differ markedly from an airy disk. This happens because only the fraction of the incident field with high incidence angles is focused, causing that the $\hat{z}$ component of the focused field has an amplitude comparable to the lateral components. For an incident beam with linear polarization along $\hat{x}$, the focus is mainly composed of the x- and z-components leading to an intensity distribution with two maxima (Figures 16a and 16c), as it has been observed experimentally[16,17]. For an incident beam with circular polarization, the three components are comparable, and a symmetric intensity distribution is obtained (Figures 16b and 16d). For both polarizations, the intensity of the focus along the axial direction decays exponentially from the interface into the sample, showing the evanescent nature of the TIR foci.

PyFocus also enables the calculation of toroidal foci in total internal reflection. In this case, the incident field was defined by multiplying a Gaussian beam modulated by a VP mask with the obstruction function: $\boldsymbol{E}_i(\rho', \phi') = \boldsymbol{E}_G(\rho')\, e^{i\phi'}\, g_{TIRF}$, which in PyFocus can be simulated using the

`custom` function and defining `entrance_field=np.exp(-(rho/w0)**2)` and `custom_mask` as:

```
theta_crit=np.arcsin(n[1]/n[0])
def custom_mask(rho,phi,w0,f,k):
    if rho>f*np.sin(theta_crit):
        return np.exp(1j*phi)
    else:
        return 0
```

The intensity distribution of the obtained focus is shown in Figures 17a-c. This simulation required an average computing time of 12 minutes.

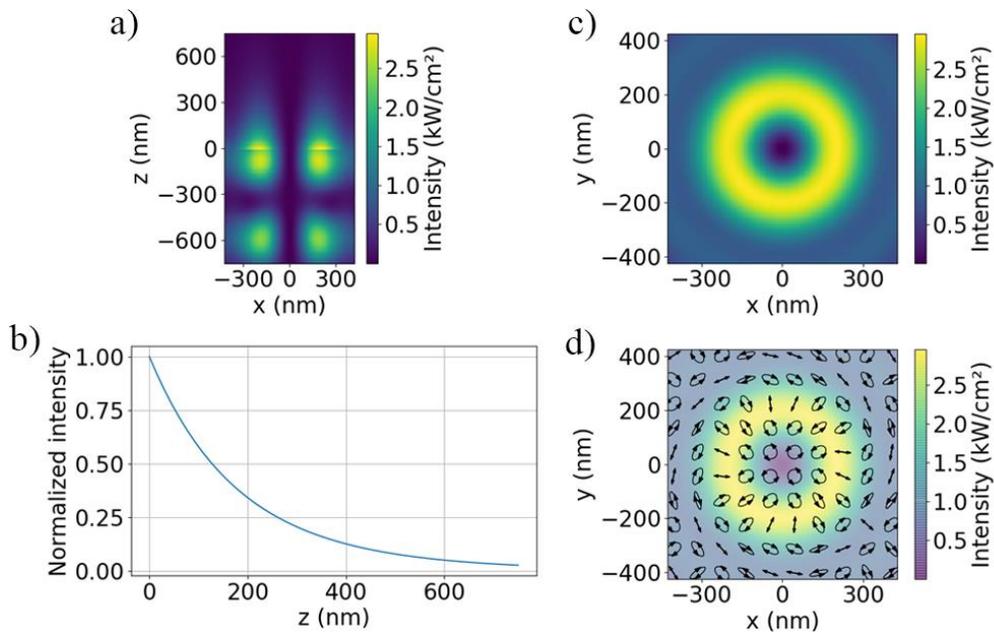

**Figure 17.** Toroidal foci in total internal reflection. a, b) Intensity on the $y = 0$ $plane$ and the $z = 5$ nm plane respectively. c) Intensity of the transmitted field on the $x = 200$ nm, $y = 0$ axis. d) Polarization map corresponding to b).

As observed previously, the intensity on the y = 0 plane shows an exponential decay along the axial direction given by the evanescent waves. On the focal plane, the toroidal shape is maintained, and the polarization remains circular at small distances from the center of the pattern (less than 150 nm, Figure 17d).

Another example of the use of a custom mask, showing a transition between the foci obtained by the full incident field and the foci generated by total internal reflection, and an example of the simulation of a multilayer system are shown in Appendix C.

# 6. Conclusions

We have then presented PyFocus, an open-source Python software package to perform fully vectorial calculations of focused electromagnetic fields after modulation by an arbitrary phase mask. We describe its basic working principles and provide application examples using functions on custom scripts or through the provided graphical user interface.

To demonstrate the potential of PyFocus, we performed an extensive characterizations of toroidal foci under various conditions relevant to experiments. Toroidal foci, also known as doughnut-shaped foci, are widely used in super-resolution fluorescence microscopy methods such as STED, RESOLFT and MINFLUX. The correct characterization of such beams is key to optimize experimental designs and to interpret properly their results. We provide examples of the influence of different experimental factors such as polarization, aberrations and misalignments of optical elements, on the resulting toroidal foci. Furthermore, we performed calculations of both gaussian and toroidal foci through an interface of different mediums. Such an experimental situation is typical in optical microscopy where the refractive index of the sample and the objective medium are often different. Also, to our knowledge, we provide the first theoretical demonstration that toroidal can be produced in total internal reflection conditions. Such foci are of interest for example in MINFLUX to increase the signal-to-background ratio of the measurements or in STED and RESOLFT to decrease the dose of light delivered to the biological sample.

Finally, we believe that PyFocus will find applications in many different fields of optics, especially where relatively high numerical apertures are involved, and full vectorial calculations are needed.

**Appendix A. Derivations**

*A.1. Focus of a Gaussian beam without phase modulation*

Equations 2 and 3 have an analytical solution by using of the relation:

$$\int_0^{2\pi} e^{im\phi'} e^{i\epsilon \cos(\phi'-\phi)} d\phi' = 2\pi i^m e^{im\phi} J_m(\epsilon) \quad (12)$$

where $J_m$ is a Bessel function of the first kind of order $m$. The field obtained by focusing a Gaussian beam then results:

$$E_{f_x} = \frac{-ikf\,A}{2}\{c_x[I_0(\rho) + I_2(\rho)\cos(2\phi)] - c_y[I_2(\rho)\sin(2\phi)]\}$$

$$E_{f_y} = \frac{-ikf\,A}{2}\{c_x[I_2(\rho)\sin(2\phi)] - c_y[I_0(\rho) - I_2(\rho)\cos(2\phi)]\}$$

$$E_{f_z} = -ikf\,AI_1(\rho)[c_x\cos(\phi) - c_y\sin(\phi)] \quad (13)$$

With $I_0, I_1, I_2$ the integrals:

$$I_0(\rho) = \int_0^\alpha d\theta\, e^{-\left(\frac{f\sin(\theta)}{w_0}\right)^2} \sqrt{\cos(\theta)}\, \sin(\theta)(1 + \cos(\theta))\, J_0(k\rho\,\sin(\theta))\, e^{ikz\cos(\theta)}$$

$$I_1(\rho) = \int_0^\alpha d\theta\, e^{-\left(\frac{f\sin(\theta)}{w_0}\right)^2} \sqrt{\cos(\theta)}\, \sin^2(\theta)\, J_1(k\rho\,\sin(\theta))\, e^{ikz\cos(\theta)}$$

$$I_2(\rho) = \int_0^\alpha d\theta \, e^{-\left(\frac{f\sin(\theta)}{w_0}\right)^2} \sqrt{\cos(\theta)} \sin(\theta)(1-\cos(\theta)) J_2(k\rho\sin(\theta)) e^{ikz\cos(\theta)} \quad (14)$$

*A.2. Focus of a Gaussian beam modulated by a VP phase mask*

Equations 2 and 3 for a Gaussian beam modulated by a VP mask, and considering the propagation between the phase mask and the objective lens, have an analytical solution using equation 10:

$$E_{f_x} = \frac{kfA}{2} \left\{ c_x \left[ I'_1(\rho)e^{i\phi} - \frac{1}{2}I'_2(\rho)e^{-i\phi} + \frac{1}{2}I'_3(\rho)e^{3i\phi} \right] - ic_y \left[ \frac{1}{2}I'_2(\rho)e^{-i\phi} + \frac{1}{2}I'_3(\rho)e^{3i\phi} \right] \right\}$$

$$E_{f_y} = \frac{kfA}{2} \left\{ ic_x \left[ \frac{-1}{2}I'_2(\rho)e^{-i\phi} + \frac{1}{2}I'_3(\rho)e^{3i\phi} \right] - c_y \left[ I'_1(\rho)e^{i\phi} + \frac{1}{2}I'_2(\rho)e^{-i\phi} - \frac{1}{2}I'_3(\rho)e^{3i\phi} \right] \right\}$$

$$E_{f_z} = \frac{kfA}{2} \left\{ ic_x [I'_4(\rho) - I'_5(\rho)e^{2i\phi}] - c_y [I'_4(\rho) + I'_5(\rho)e^{2i\phi}] \right\} \quad (15)$$

with $I'_1, I'_2, I'_3, I'_4, I'_5$ the integrals:

$$I'_1(\rho) = \frac{k}{L} \int_0^\alpha d\theta \, e^{ik\left(\frac{\rho'^2}{2L}\right)} I(f\sin(\theta)) \sqrt{\cos(\theta)} \sin(\theta) J_1(k\rho\sin(\theta))(1+\cos(\theta)) e^{ikz\cos(\theta)}$$

$$I'_2(\rho) = \frac{k}{L} \int_0^\alpha d\theta \, e^{ik\left(\frac{\rho'^2}{2L}\right)} I(f\sin(\theta)) \sqrt{\cos(\theta)} \sin(\theta) J_1(k\rho\sin(\theta))(1-\cos(\theta)) e^{ikz\cos(\theta)}$$

$$I'_3(\rho) = \frac{k}{L} \int_0^\alpha d\theta \, e^{ik\left(\frac{\rho'^2}{2L}\right)} I(f\sin(\theta)) \sqrt{\cos(\theta)} \sin(\theta) J_3(k\rho\sin(\theta))(1-\cos(\theta)) e^{ikz\cos(\theta)}$$

$$I'_4(\rho) = \frac{k}{L} \int_0^\alpha d\theta \, e^{ik\left(\frac{\rho'^2}{2L}\right)} I(f\sin(\theta)) \sqrt{\cos(\theta)} \sin^2(\theta) J_0(k\rho\sin(\theta)) e^{ikz\cos(\theta)}$$

$$I'_5(\rho) = \frac{k}{L} \int_0^\alpha d\theta \, e^{ik\left(\frac{\rho'^2}{2L}\right)} I(f\sin(\theta)) \sqrt{\cos(\theta)} \sin^2(\theta) J_2(k\rho\sin(\theta)) e^{ikz\cos(\theta)} \quad (16)$$

and $I(f\sin(\theta)) = \int_0^R \rho'' d\rho''\, e^{-\left(\frac{\rho''}{w_0}\right)^2} e^{ik\left(\frac{\rho''^2}{2L}\right)} J_1(k\rho'' f\sin(\theta)/L)$.

If the propagation between the phase mask and the objective lens is depreciated, these integrals are given by:

$$I'_1(\rho) = \frac{k}{L}\int_0^\alpha d\theta\, e^{-\left(\frac{f\sin(\theta)}{w_0}\right)^2} \sqrt{\cos(\theta)}\sin(\theta) J_1(k\rho\sin(\theta))(1+\cos(\theta))\, e^{ikz\cos(\theta)}$$

$$I'_2(\rho) = \frac{k}{L}\int_0^\alpha d\theta\, e^{-\left(\frac{f\sin(\theta)}{w_0}\right)^2} \sqrt{\cos(\theta)}\sin(\theta) J_1(k\rho\sin(\theta))(1-\cos(\theta))\, e^{ikz\cos(\theta)}$$

$$I'_3(\rho) = \frac{k}{L}\int_0^\alpha d\theta\, e^{-\left(\frac{f\sin(\theta)}{w_0}\right)^2} \sqrt{\cos(\theta)}\sin(\theta) J_3(k\rho\sin(\theta))(1-\cos(\theta))\, e^{ikz\cos(\theta)}$$

$$I'_4(\rho) = \frac{k}{L}\int_0^\alpha d\theta\, e^{-\left(\frac{f\sin(\theta)}{w_0}\right)^2} \sqrt{\cos(\theta)}\sin^2(\theta) J_0(k\rho\sin(\theta))\, e^{ikz\cos(\theta)}$$

$$I'_5(\rho) = \frac{k}{L}\int_0^\alpha d\theta\, e^{-\left(\frac{f\sin(\theta)}{w_0}\right)^2} \sqrt{\cos(\theta)}\sin^2(\theta) J_2(k\rho\sin(\theta))\, e^{ikz\cos(\theta)} \qquad (17)$$

*A.3. Focused fields at each side of a multilayer system*

When the focusing field $\boldsymbol{E}_0$ travels through a multilayer system, the field may take different forms at each side of the multilayer. In particular, for an incident field with elliptical polarization it can then be expressed as:

$$\boldsymbol{E}_f = \begin{cases} c_x \boldsymbol{E}_{f0}^{(x)} + c_x \boldsymbol{E}_r^{(x)} + c_y \boldsymbol{E}_{f0}^{(y)} + c_y \boldsymbol{E}_r^{(y)} & \text{if } z < z_{int} \\ c_x \boldsymbol{E}_t^{(x)} + c_y \boldsymbol{E}_t^{(y)} & \text{if } z > z_{int} \end{cases} \qquad (18)$$

Where $\boldsymbol{E}_{f0}^{(x)}$ and $\boldsymbol{E}_{f0}^{(y)}$ are the focused fields in the absence of the multilayer system produced by incident fields linearly polarized along $\hat{\boldsymbol{x}}$ and $\hat{\boldsymbol{y}}$, respectively. $\boldsymbol{E}_r$ and $\boldsymbol{E}_t$ are the fields reflected and transmitted by the multilayer, respectively, can be expressed as[1]:

$$\boldsymbol{E}_r = \frac{-if\, e^{-ik_1 f}}{2\pi} \iint_{k_x,k_y} \boldsymbol{E}_{0r} \frac{1}{k_{z_1}} e^{i(k_x x + k_y y - k_{z_1} z)} \sqrt{\cos(\theta)}\, dk_x dk_y$$

$$\boldsymbol{E}_t = \frac{-if\, e^{-ik_1 f}}{2\pi} \iint_{k_x,k_y} \boldsymbol{E}_{0t} \frac{1}{k_{z_2}} e^{i(k_x x + k_y y + k_{z_2} z)} \sqrt{\cos(\theta)}\, dk_x dk_y \qquad (19)$$

Where equation 19 is expressed in terms of $k_x$ and $k_y$, instead of $\theta$ and $\phi'$. This expression in terms of $k$ is preferable since it allows expressing $\boldsymbol{E}_t$ without employing the inclination angle at the exit of the multilayer as an integration variable for the field obtained by evanescent waves, since for this part of the field it has a complex value. $k_1$ and $k_2$ are the wavenumbers in the medium of the lens and in the medium of the sample, respectively. While $k_x$ and $k_y$ remain unaltered through the multilayer, the $\hat{z}$ component is defined as $k_{z_1} = \sqrt{k_1^2 - k_x^2 - k_y^2}$ previous to the interface an $k_{z_2} = \sqrt{k_2^2 - k_x^2 - k_y^2}$ after the interface.

$\boldsymbol{E}_{0r}$ and $\boldsymbol{E}_{0t}$ are the reflected and transmitted components of $\boldsymbol{E}_0$, which can be expressed in terms of $\boldsymbol{E}_i$ and the reflection and transmission coefficients for the $s$ and $p$ components ($r^s$, $r^p$, $t^s$ and $t^p$) Expressing the integrals in equation 16 using the spherical variables ($\theta$, $\phi'$), the focused fields at each side of the multilayer system, for an incident field linearly polarized along $\hat{x}$ result:

$$\boldsymbol{E}_r^{(x)} = -\frac{ik_1 f e^{-i(k_1 f + 2k_{z_1} z_{int})}}{2\pi} \int_0^{2\pi} \int_0^{\alpha} e^{2i k_1 \cos(\theta) z_{int}} \boldsymbol{E}_i(\theta, \phi') \cdot$$

$$\cdot \begin{bmatrix} r^s(\theta)\sin^2(\phi') - r^p(\theta)\cos^2(\phi')\cos(\theta) \\ -(r^s(\theta) + r^p(\theta)\cos(\theta))\cos(\phi')\sin(\phi') \\ -r^p(\theta)\sin(\theta)\cos(\phi') \end{bmatrix} e^{ik_1(-z\cos(\theta) + \rho\sin(\theta)\cos(\phi'-\phi))} \sqrt{\cos(\theta)}\sin(\theta) d\theta d\phi'$$

$$\boldsymbol{E}_t^{(x)} = -\frac{ik_1 f e^{-i[k_1 f + 2(k_{z_1} - k_{z_2}) z_{int}]}}{2\pi} \int_0^{2\pi} \int_0^{\alpha} e^{i[k_1\cos(\theta) - k_2\cos(\theta_t)] z_{int}} \boldsymbol{E}_i(\theta, \phi') \cdot$$

$$\cdot \begin{bmatrix} t^s(\theta)\sin^2(\phi') + t^p(\theta)\cos^2(\phi')\cos(\theta_t) \\ (-t^s(\theta) + t^p(\theta)\cos(\theta_t))\cos(\phi')\sin(\phi') \\ -t^p(\theta)\sin(\theta_t)\cos(\phi') \end{bmatrix} e^{ik_2(-z\cos(\theta_t) + \rho\sin(\theta_t)\cos(\phi'-\phi))} \sqrt{\cos(\theta)}\sin(\theta) d\theta d\phi' (20)$$

Where $\theta_t$ is the inclination angle at the exit of the multilayer, such that $cos(\theta_t) = \sqrt{1 - \frac{n_1^2}{n_2^2}\sin^2(\theta)}$ and $\sin(\theta_t) = \frac{n_1}{n_2}\sin(\theta)$. As described in section 2.1, with the use of coordinate rotation the field obtained for an incident field with linear polarization along $\hat{y}$ results:

$$\boldsymbol{E}_r^{(y)} = -\frac{ik_1 f e^{-i(k_1 f + 2k_{z1} z_{int})}}{2\pi} \int_0^{2\pi} \int_0^{\alpha} E_i(\theta, \phi') \begin{bmatrix} (r^s(\theta) + r^p(\theta)\cos(\theta))\cos(\phi')\sin(\phi') \\ r^s(\theta)\sin^2(\phi') - r^p(\theta)\cos^2(\phi')\cos(\theta) \\ -r^p(\theta)\sin(\theta)\cos(\phi') \end{bmatrix} \cdot$$

$$\cdot e^{ik_1(-z\cos(\theta) - \rho \sin(\theta)\sin(\phi' - \phi))} \sqrt{\cos(\theta)} \sin(\theta) d\theta d\phi'$$

$$\boldsymbol{E}_t^{(y)} = -\frac{ik_1 f e^{-i[k_1 f + 2(k_{z1} - k_{z2})z_{int}]}}{2\pi} \int_0^{2\pi} \int_0^{\alpha} E_i(\theta, \phi') \begin{bmatrix} (t^s(\theta) - t^p(\theta)\cos(\theta_t))\cos(\phi')\sin(\phi') \\ t^s(\theta)\sin^2(\phi') + t^p(\theta)\cos^2(\phi')\cos(\theta_t) \\ -t^p(\theta)\sin(\theta_t)\cos(\phi') \end{bmatrix} \cdot$$

$$\cdot e^{ik_2(-z\cos(\theta_t) - \rho \sin(\theta_t)\sin(\phi' - \phi))} \sqrt{\cos(\theta)} \sin(\theta) d\theta d\phi' \qquad (21)$$

*A.4. Focusing through a positive-hand VP mask with a left circular polarization beam*

In Figure 18 we show the field obtained by focusing a Gaussian beam of left circular polarization modulated by a positive-hand VP phase mask with the use of the `VP` function (`fields=sim.VP(propagation=False, multilayer=False,*parameters)` while setting the polarization with `gamma=45`, `beta=-90`). In Figures 18a and 18b it can be seen that the formed focus does not present a central zero. Inspection of the different components (Figures 18c-e) show that the intensity of the $\hat{z}$ component has an intensity maximum at the center, which avoids the expected toroidal shape.

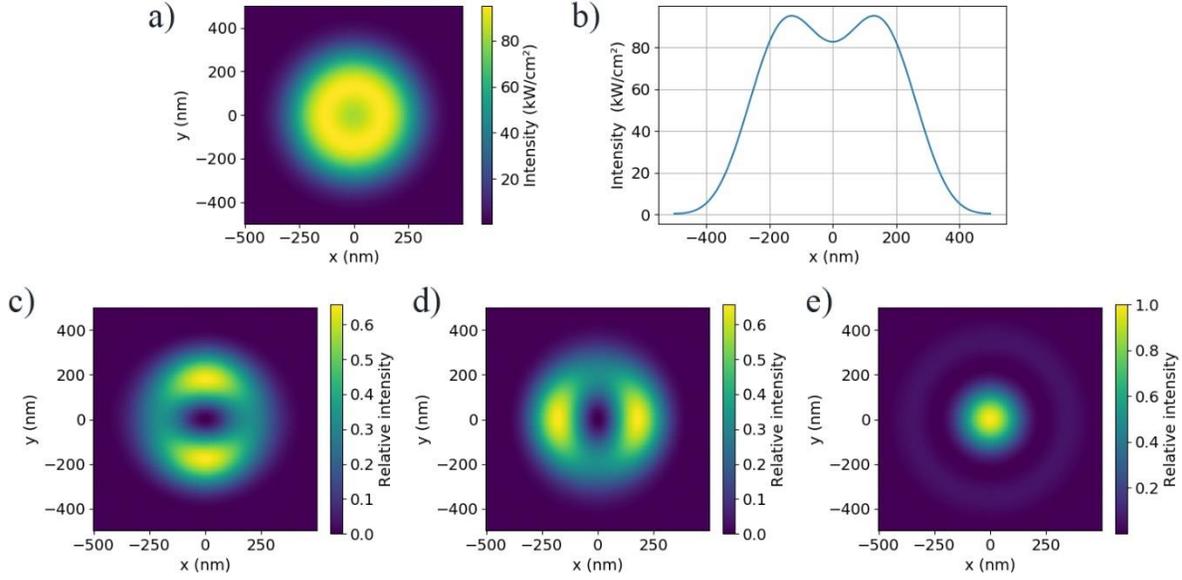

**Figure 18.** Field obtained by focusing a Gaussian beam of left circular polarization modulated by a VP phase mask. a, b) Intensity on the $z = 0$ plane and the $z = 0$, $y = 0$ axis. c, d, e) Intensity of the cartesian components on the $z = 0$ plane.

## Appendix B. Technical details

### B.1. Custom incident fields using a user given .txt file

If the incident field is given by a `.txt` file, this file must be composed of a 2D array with the complex amplitude of the incident field at the $(\phi', \theta)$ points used in the 2D numerical integration. Each row should denote the $\phi'$ coordinate in an array `np.linspace(0, 2π, N_φ)` and each column should denote the $\theta$ coordinate in an array `np.linspace(0, α, N_θ)`, with $N_\phi, N_\theta$ two integers defining the resolution of the calculus.

### B.2. Number of divisions for 2D numerical integrations

To evaluate the number of divisions in the 2D numerical integrations, the toroidal foci obtained by using the `custom` function and defining the incident field function as a gaussian beam modulated by a VP mask: `entrance_field = lambda rho,phi,w0,f,k: np.exp(-(rho/w0)**2)`, `custom_mask= np.exp(1j*phi)`, with different amounts of divisions for the integration coordinates, $N_\phi$ for the coordinate $\phi'$ and $N_\theta$ for $\theta$, were simulated. The obtained foci for $N_\theta = N_\phi = 10$, $N_\theta = N_\phi = 100$ and $N_\theta = N_\phi = 200$ are shown in Figure 19. For $N_\theta = N_\phi = 10$, the expected circular symmetry is not found, indicating that the resolution is not sufficient. For $N_\theta = N_\phi = 100$ and $N_\phi = N_\theta = 200$, the expected symmetry is obtained, and no appreciable differences can be observed between the two simulations. The same analysis was performed for all the incident fields used in our examples, finding that $N_\phi = N_\theta = 200$ is a value high enough to not show appreciable errors in the simulations while maintaining average computing times of no more than tens of minutes.

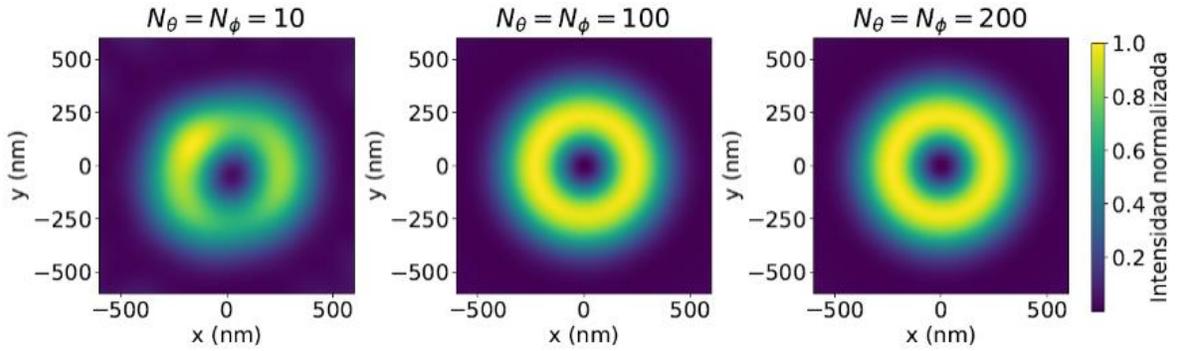

**Figure 19.** Obtained toroidal foci using the `custom` function for various numbers of divisions $N_\theta$ and $N_\phi$.

**Appendix C. Further examples of custom masks**

*C.1. Transition from full illumination to TIRF*

To show a transition between the foci obtained by the full incident field and the foci generated by total internal reflection, we calculated the field obtained with the use of five obstruction functions $g_i$:

$g_1(\rho') = 1$, $g_2(\rho') = \frac{\rho'}{2h} + \frac{1}{2}$, $g_3(\rho') = \frac{\rho'}{h}$, $g_4(\rho') = \begin{cases} \frac{\rho'-h/2}{h/2} & \text{if } \rho' > h/2 \\ 0 & \text{if } \rho' < h/2 \end{cases}$, $g_5(\rho') = g_{TIRF}(\rho') = \begin{cases} 1 \text{ if } \rho' > f\sin(\theta_c) \\ 0 \text{ if } \rho' < f\sin(\theta_c) \end{cases}$.

A graphic of these functions is shown in Figure 20a, and the foci obtained for each function and for an interface located at a distance $z_{int} = -200$nm from the focal plane is shown in Figure 20b. Each simulation required an average computing time of 12 minutes.

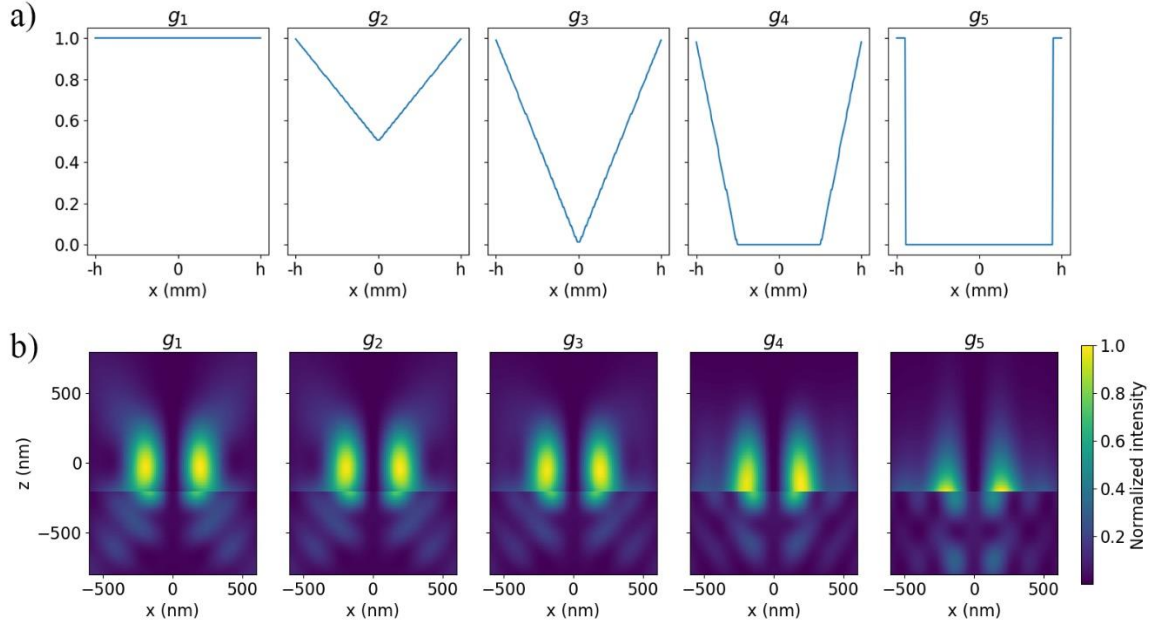

**Figure 20.** Toroidal foci obtained for various obstruction functions $g_i$ for a glass-water interface located at a distance $z_{int} = -200$ nm from the focal plane. a) Graphic of each of the obstruction functions. b) Intensity on the $y = 0$ plane. As the portion of the incident field that does not generate evanescent waves is blocked the field penetrates less on the axial direction.

It can be seen that as the portion of the incident field that does not generate evanescent waves is blocked the field penetrates less on the axial direction.

*C.2. Field through a glass-gold-spacer-water interface*

To show the simulation of a field obtained through a multilayer system, we simulated the case of a glass-gold-spacer-water interface. This multilayer was evaluated in another work[18] and allows a further validation for the code. The used refraction indexes were $1.5, 0.14 + 3.55j, 1.54, 1.33$ (in the core code this results in `n  =  np.array((1.5,0.14+3.55j,1.54,1.33))`) and the
38

thickness was set to 44, 24 (in the core code this results in `ds = np.array((np.inf,44,24, np.inf))` since the glass and water mediums are considered semi-infinite). The amplitude of transmission coefficient of the p component of this multilayer system is shown in Figure 21a. The multilayer system was set at the focal plane ($z_{int} = 0$). The resulting transmitted field for a Gaussian beam of linear polarization along $\hat{x}$ without phase modulation is shown in Figures 21b and 21c. This simulation required an average computing time of 12 minutes.

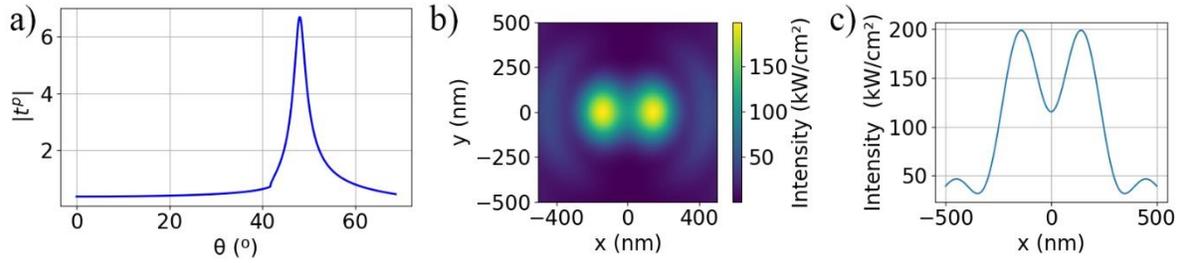

**Figure 21.** Foci obtained by focusing a Gaussian beam of linear polarization along $\hat{x}$ through a multilayer system made of glass ($n = 1.5$), 44 nm gold ($n = 0.14 + 3.55j$), polymer ($n = 1.54$), and water ($n = 1.33$). a) Transmission amplitude of the p component as a function of the incidence angle. b) Focused field intensity on the $z = 5$ nm plane. c) Intensity on the $z = 5$ nm, $y = 0$ axis.

The obtained pattern differs markedly from an airy disk, presenting two maxima because the $\hat{z}$ component of the transmitted field is of comparable amplitude to the lateral components[19].

# Declaration of interests

The authors declare that they have no known competing financial interests or personal relationships that could have appeared to influence the work reported in this paper.

# Author contributions

**Fernando Caprile:** Software, Resources, Investigation, Visualization, Writing – Original Draft.

**Luciano A. Masullo:** Conceptualization, Methodology, Validation, Writing – Original Draft, Supervision

**Fernando D. Stefani:** Conceptualization, Methodology, Validation, Writing – Review and Editing, Supervision, Project Administration, Funding Acquisition

**FUNDING**

This work has been funded by CONICET, ANPCYT Projects PICT-2017-0870, and PICT-2014-0739.

**ACKNOWLEDGMENTS**

F.D.S. acknowledges the support of the Max-Planck-Society and the Alexander von Humboldt Foundation.